\documentclass[a4paper]{article}
\usepackage{amsfonts}
\usepackage{amsmath}
\usepackage{amssymb}
\usepackage[a4paper,includeheadfoot]{geometry}

\input{tcilatex}

\begin{document}

\title{Renormalization: The observable-state model}
\author{Juan Sebasti\'{a}n Ardenghi \\
Instituto de Astronom\'{\i}a y F\'{\i}sica del Espacio. \and Mario
Castagnino \\
Institutos de F\'{\i}sica de Rosario y de Astronom\'{\i}a y F\'{\i}sica del
Espacio.\\
Casilla de Correos 67, sucursal 28, 1428 Buenos Aires, Argentina.}
\maketitle

\begin{abstract}
The usual mathematical formalism of quantum field theory is not rigorous
because it contains divergences that can only be renormalized by
non-rigorous mathematical methods. So we present a method of subtraction of
divergences using the formalism of decoherence. This is achieved by
replacing the standard renormalization method by a projector on a well
defined Hilbert subspace. In this way a list of problems of the standard
formalism disappears while the physical results of QFT remain valid. From
its own nature, this formalism can be also used in non-renormalizable
theories.
\end{abstract}

\section{Introduction}

The development of formalisms encompassing several chapters of physics is
one of the main purposes of theoretical physics. Experience shows that when
two chapters are successfully unified, the obtained formalism frequently
explains new phenomena which were not included in neither one of the two
chapters: the unification of electrostatic and magnetism being a venerable
and eloquent example. The basis of a unification is the choice of a common
mathematical structure, e. g. many physical systems share a common feature:
only some part of the information they contain is relevant. Following this
line, in this paper, we present a common formalism for some features of
decoherence and Quantum Field Theory (QFT), two theories that deal with this
kind of systems, with the result that some new bursts of light are seeded in
the former theory.

The comprehension of both decoherence theory and QFT was greatly improved in
the last decades. Morever, now a days we understand the mechanics of
decoherence and the classical limit quite well. Nevertheless, there is not
an accepted rigorous formalism of QFT, because many doubts still remain. In
fact, QFT has a certain bad reputation: mathematicians say that it is not
properly formulated, philosophers find that some old unsolved problems
reappear in QFT in a virulent shape,\footnote{%
Like the one of internal and external relations \ (\cite{Sunny}, page 190).}
and some physicists feel that something is not completely clear.\footnote{%
Many years ago K. O. Friedrichs said: \textquotedblleft Quantum Field Theory
is akin to the challenge felt by an archeologist stumbling on records of a
high civilization written in strange symbols. Clearly there were intelligent
messages but what did they want to say?\textquotedblright\ (Even if the
sentence is old it is still standing since Haag quoted it in his book \cite%
{Haag}). P. Roman also said that in QFT we have only learned to
\textquotedblleft peacefully coexist\textquotedblright\ with alarming
divergencies (\cite{Roman}, page. 298). P. Ramond (\cite{Ramond} page 172)
and L. S. Brown considered the renormalization a \textquotedblleft
miracle\textquotedblright\ (\cite{Brown}, page 243), etc. (see also \cite%
{Teller} and \cite{Sunny}).
\par
Of course this is not a universal opinion and may be an extreme one, but it
is certainly the one, e. g., of Haag's. This will be the point of view that
we will adopt in this paper, even if we acknowledge other most respectable
opinions, e.g.: the explanation of renormalization based in an analogy with
statistical physics of magnets and fluids \cite{PS}.} For these reasons
alternative theories were developed: the axiomatic version, superstrings,
branes, loop quantum gravity, etc (see \cite{Pol}, \cite{kiefer}) . This
paper is an attempt to explain QFT using another approach based on several
ideas, mainly the proper definition of quantum states and observables and
new techniques to deal with systems with continuous evolution spectrum,
which gave good results in other cases (\cite{castagnino}, \cite{C2}, \cite%
{Deco}, \cite{4'}, \cite{4-2}, \cite{C3}).\footnote{%
The continuous spectrum will force us to work with distributions, kernels,
etc. We will do so, instead of putting the system in a box, lattice, etc. In
this way we will obtain a more direct explanation of really what is going on.%
}

\subsection{The two main ideas.}

The main purpose of this paper is to show the equivalence between the
quantum theory of fields of $\phi ^{4}$, and what we will call the
mathematical formalism for quantum continuous systems that will be
introduced in section 3. Following the main idea of \cite{I}, in this work
it will be shown that the generating functional of $\phi ^{4}$ theory can be
written as the sum of two terms: a divergent term, which contains all the
infinities of the theory, and a regular term which contains the physical
contribution.

Our program is based on the introduction of a rigorous mathematical
formalism based in two main ideas:

1.- We will deal with quantum systems where partial degrees of freedom are
used and other partial degrees of freedom are neglected. In QFT, the
counterterms of renormalization theory eliminate some part of information
that it is considered unphysical since it contains meaningless infinities.
Analogously, in the formalism introduced in this work, the whole quantum
system is descomposed in an external quantum system and in an internal
quantum system, but only the relevant degrees of freedom will be considered
and these will correspond with the degrees of freedom of the external
quantum system.

2.- We will substitute the unsatisfactory counterterms in QFT
renormalization by a simple projection $\Pi $ on a well defined subspace of
an also well defined Hilbert space. The central idea is the following: if $%
\tau ^{(n)}(x_{1},...,x_{n})$ are some (symmetric) $n$-point functions (like
Feynman or Euclidean functions) we can define the corresponding generating
functional (\cite{Haag}, eq. (II.2.21), \cite{Brown}, eq. (3.2.11)) as:

\begin{equation}
W\left[ J\right] =\underset{n=0}{\overset{\infty }{\sum }}\frac{i^{n}}{n!}%
\dint \tau ^{(n)}(x_{1},...,x_{n})J(x_{1})...J(x_{n})d^{4}x_{1}...d^{4}x_{n}
\label{ideas1}
\end{equation}%
where:\footnote{%
In a realistic field theory (a theory with interactions), the functions of
eq.(\ref{ideas2}) are badly defined, since they are objects with
mathematical properties that are worse than those of the distributions.}

\begin{equation}
\tau ^{(n)}(x_{1},...,x_{n})\sim \left\langle 0\left\vert \phi
(x_{1})...\phi (x_{N})\right\vert 0\right\rangle  \label{ideas2}
\end{equation}%
A convenient way to to eliminate trivial contributions of single-particle
propagators is by introducing a modified generating functional $Z[J]$ for
irreducible Green's functions. It is defined as

\begin{equation}
W\left[ J\right] =e^{iZ[J]}  \label{ideas3}
\end{equation}%
The new generating functional $Z[J]$ satisfies the normalization condition $%
Z[0]=0$ and it reads:

\begin{equation}
iZ\left[ J\right] =\underset{n=0}{\overset{\infty }{\sum }}\frac{i^{n}}{n!}%
\dint \tau
_{c}^{(n)}(x_{1},...,x_{n})J(x_{1})...J(x_{n})d^{4}x_{1}...d^{4}x_{n}
\label{ideas4}
\end{equation}%
where in this case $\tau _{c}^{(n)}(x_{1},...,x_{n})$ are connected n-point
functions that can be obtained by differentiation

\begin{equation}
\tau _{c}^{(n)}(x_{1},...,x_{n})=\frac{1}{i^{n-1}}\frac{\delta ^{n}Z[J]}{%
\delta J(x_{1})...\delta J(x_{n})}\mid _{J=0}  \label{ideas5}
\end{equation}%
In turn, the connected $n$-point functions can be written in terms of the
Lagrangian interaction density $\mathcal{L}_{I}^{0}(y_{p})$ as

\begin{equation}
\tau _{c}^{(n)}(x_{1},...,x_{n})^{(p)}=\frac{i^{p}}{p!}\dint \left\langle
\Omega _{0}\left\vert T\phi (x_{1})...\phi (x_{n})\mathcal{L}%
_{I}^{0}(y_{1})...\mathcal{L}_{I}^{0}(y_{p})\right\vert \Omega
_{0}\right\rangle d^{4}y_{1}...d^{4}y_{p}  \label{ideas6}
\end{equation}%
Introducing (\ref{ideas6}) in (\ref{ideas4}) we have

\begin{equation}
iZ\left[ J\right] =\underset{n=0}{\overset{\infty }{\sum }}\underset{p=0}{%
\overset{\infty }{\sum }}\frac{i^{n}}{n!}\frac{i^{p}}{p!}\dint \left\langle
\Omega _{0}\left\vert T\phi (x_{1})...\phi (x_{n})\mathcal{L}%
_{I}^{0}(y_{1})...\mathcal{L}_{I}^{0}(y_{p})\right\vert \Omega
_{0}\right\rangle
J(x_{1})...J(x_{n})d^{4}y_{1}...d^{4}y_{p}d^{4}x_{1}...d^{4}x_{n}
\label{ideas7}
\end{equation}

The main idea of this paper is to rewrite the generating functional of
connected Feynman diagrams (eq. \ref{ideas7}) as the inner product of a
state with an observable. The observables will have the property of being
diagonal in some of its components which will contain the short-distance
singularities of the physical theory. In turn, these singularities will
appear in the inner product if also the state has a non-zero diagonal part
in the same components. In this way, the physical contribution will be
obtained by throwing away the diagonal part of the state by a projection in
the Hilbert space where the state are defined.

This procedure has a conceptual counterpart: essentially we must admit that
the main role of physics is to explain what the apparatuses measure. To do
this physicists usually construct an ideal model of the system under study,
using postulates and mathematical structures that go far beyond the simple
measurements of the apparatuses (e. g. the unitary time evolution theories,
or when we only consider the microstates of a system, etc.). In fact, it is
very rare to model a physical system accurately, and so it is quite usual to
construct models which only vaguely resemble the real system but whose
essence one hopes to capture. This is the case of irreversibility and
decoherence but also the case of QFT, where the Lagrangians are usually
chosen only by their simplicity and covariant properties. But after a model
of the system is adopted, physicists again consider the apparatuses and what
they really measure and they refine the set of states only considering those
that are real and measurable. Namely, they restrain the whole information
the system ideally contains, only keeping the information that the
apparatuses really provide and rejecting the rest \cite{Jaynes} (e. g. when
they obtain non-unitary time evolution theories via coarse-graining or the
consideration of the macrostates only, etc.). Then if the theoretical
prediction coincides with the measurements up to a certain level they say
that the theory is correct (up to this level). In some theories this fact is
clearly stated (e. g. in decoherence theory, see paper \cite{10-1}), but not
in others. Following this line of thought, our presentation in QFT coincides
with the idea of restrain the whole information that the quantum field
contains. These ideas agree with those state in \cite{W} (vol. 1, page 499):
QFT yields divergent integral \textquotedblleft but these infinities cancel
when we express all the parameters of the theory in renormalized\
quantities, such as the masses and the charges \textit{that we actually
measure}\textquotedblright . Morever, it also coincides with \cite{Folland},
since we believe that the process of subtracting infinities is really a
matter of subtracting the irrelevant effect of the \textquotedblleft perhaps
poorly understood physics at high energy or short scale to obtain the
meaningful physics at the scales actually studied in the
laboratory\textquotedblright . In this sense, the restraining is done by
neglecting the physics of high energy or short scale.

In the standard presentation of QFT in textbooks, the infinities are
eliminated by the introduction of counterterms in the Lagrangian. This is a
nonaesthetic and poorly motivated method. Really the simpler BPHZ
subtraction of infinities introduced long ago in papers \cite{ref} is more
direct. We will restudy this method using dimensional regularization \cite%
{TW} and we will show that the divergences can be avoided by restraining the
quantum state of the quantum field with a projector. In this way the
substraction will not be an ad hoc procedure to make finite an essentially
divergent theory, but it will be the consequence of the projector that does
not see the short-scale behaviour of the quantum field.

Even if our mathematical treatment is essentially rigorous, in this paper we
do not intend to give an axiomatic version for philosophers nor a
mathematical development suitable for pure mathematicians' minds (these
matters are only sketched and they will be explained elsewhere). On the
contrary we will try to present \ a treatment that could be meaningful for
physicists, mathematicians and philosophers of physics. To do this we will
focus in some apparently irrelevant details to make our exposition as clear
as possible. Finally, the main advantage of this method is the possible
application to non-renormalizable theories that will be studied in future
works. For the sake of simplicity we have only added the second order in the
perturbation expansion for the self-energy of the electron in $QED$ and the
first order in the perturbation expansion for $\phi ^{6}$ theory.

In section 2 we will study the decoherence phenomenon in the discrete and
continuous case showing how divergences naturally appear in the later case.
In section 3 we will introduce the divergent and regular structure of the
continuous quantum systems. In section 4 we will study the first order in
perturbation $\phi ^{4}$ theory to explain carefully the relation between
QFT and the continuous quantum systems. In section 5 we will study how we
can proceed with all orders in perturbation $\phi ^{4}$ theory. In section 6
we will give a conceptual explanation of the projection in algebraic terms.%
\footnote{%
In this section we will return to the concept of instrument and system model.%
} The conclusions will be stated in section 7. In Appendix A we will
calculate the number of ultraviolet divergences in a $\phi ^{l}$ theory. In
appendix B we introduce the mass shift in the two-point correlation function
in $\phi ^{4}$ theory using dimensional regularization. Finally, in Appendix
C and D we introduce how to apply the formalism introduced in Section 3 to $%
QED$ and $\phi ^{6}$ theory.

\section{Decoherence.}

\subsection{The formalism in the discrete case}

In general, to obtain irreversibility and decoherence, only some (relevant)
information must be considered, while the remaining (irrelevant) information
must be forgotten. This is the case for all the formalisms of decoherence,
including the Environment Induced Decoherence (EID) (see e. g. \cite{JPZ})
and our formalism for decoherence (SID), (that was introduced and studied in
papers \cite{castagnino}, \cite{C2}, \cite{Deco}, \cite{C3}, and \cite%
{Ordoniez}). Both formalisms are based in a choice of a space of relevant
observables and in both cases a projector $\Pi $ can be defined (see \cite%
{Fortin}). To give an example of projector in decoherence theory we will
only consider the paradigmatic EID formalism.\ In EID, a system $S$ (usually
a small system of macroscopic nature) and an environment $E$ (usually a big
system of microscopic nature) are defined (in a more or less arbitrary way)
and the closed system$\ U$ \textquotedblleft the universe\textquotedblright\
becomes $U=E\cup S.$ Then we have the system and environment subspaces $%
\mathcal{O}_{E}$ and $\mathcal{O}_{S}$ and the observable space $\mathcal{O}%
_{U}$ such that 
\begin{equation}
\mathcal{O}_{U}=\mathcal{O}_{S}\otimes \mathcal{O}_{E}  \label{EIDgen-01}
\end{equation}%
Then we consider the relevant observables $O_{R}$ defined as%
\begin{equation}
O_{R}=O_{S}\otimes I_{E}  \label{EIDgen-02}
\end{equation}%
where $O_{S}\in \mathcal{O}_{S}$ and $I_{E}$ is the identity observable of $%
\mathcal{O}_{E}$. As $U=E\cup S$ the corresponding Hilbert space is $%
\mathcal{H}_{U}=\mathcal{H}_{S}\otimes \mathcal{H}_{E}.$ Let $\{|i\rangle \}$
$(i=1,2,...,m)$ be the basis of $\mathcal{H}_{S},$ let $\{|\alpha \rangle \}$
$(\alpha =1,2,...,n)$ be the basis of $\mathcal{H}_{E}$, therefore $%
\{|i,\alpha \rangle \}$ is the basis of $\mathcal{H}_{U}$. Under these
conditions we are only interested in what the relevant observable sees, i.
e. in the mean values: 
\begin{equation}
\langle O_{R}\rangle _{\rho }=\sum_{ij\alpha \beta }\rho _{i\alpha ,j\beta
}O_{S\text{ }ij}\delta _{\alpha \beta }=\sum_{ij}\left( \sum_{\alpha }\rho
_{i\alpha ,j\alpha }\right) O_{ij}=\langle O_{S}\rangle _{\rho _{S}}
\label{EIDgen-03}
\end{equation}%
where it can easily be proved that%
\begin{equation}
\rho _{S}=Tr_{E}\rho =\sum_{\alpha }\rho _{i\alpha ,j\alpha }
\label{EIDgen-04}
\end{equation}%
where $Tr_{E}$ is the partial trace of the indices $\alpha $ of the
enviroment. In many cases it can be proved that this $\rho _{S}(t)$ evolves
in a non unitary way and reaches equilibrium at a relaxation time $t_{R}$.
Moreover a moving preferred basis can be defined where $\rho _{S}(t)$
becomes diagonal in a decoherence time $t_{D}<t_{R}$ (see \cite{Bonachon}).

\subsection{The formalism in the continuous case}

In this case, the corresponding Hilbert space is $\mathcal{H}_{U}=\mathcal{H}%
_{S}\otimes \mathcal{H}_{E}$ where $\{|\omega _{S}\rangle \}$ ($\omega
_{s}\in 
\mathbb{R}
$) is the basis of $\mathcal{H}_{S},$ and $\{|\omega _{E}\rangle \}$ ($%
\omega _{E}\in 
\mathbb{R}
$) is the basis of $\mathcal{H}_{E}$, therefore $\{|\omega _{S},\omega
_{E}\rangle \}$ is the basis of $\mathcal{H}_{U}$. If we consider the
relevant observables $O_{R}$ (see eq.(\ref{EIDgen-02})), the mean value can
be calculated as:

\begin{eqnarray}
\langle O_{R}\rangle _{\rho } &=&\dint \dint \dint \dint \rho (\omega
_{S},\omega _{E},\omega _{S}^{\prime },\omega _{E}^{\prime })O(\omega
_{S},\omega _{S}^{\prime })\delta (\omega _{E}-\omega _{E}^{\prime })d\omega
_{S}d\omega _{E}d\omega _{S}^{\prime }d\omega _{E}^{\prime }=  \label{c1} \\
&=&\dint \dint \left( \dint \rho (\omega _{S},\omega _{E},\omega
_{S}^{\prime },\omega _{E})d\omega _{E}\right) O(\omega _{S},\omega
_{S}^{\prime })d\omega _{S}d\omega _{S}^{\prime }=\langle O_{S}\rangle
_{\rho _{S}}  \notag
\end{eqnarray}%
where%
\begin{equation}
\rho _{S}=Tr_{E}\rho =\dint \rho (\omega _{S},\omega _{E},\omega
_{S}^{\prime },\omega _{E})d\omega _{E}  \label{c2}
\end{equation}%
which is the equivalent to eq.(\ref{EIDgen-04})\ in the continuous case.

\subsubsection{Divergences in the continuous formalism}

For the sake of simplicity we will only consider an isolated quantum system
with corresponding Hilbert space $\mathcal{H}$ and a basis $\left\{
\left\vert \omega \right\rangle \right\} $. The relevant observables acting
in $\mathcal{H}\otimes \mathcal{H}$ are:

\begin{equation}
O=\dint \dint (O_{D}(\omega )\delta (\omega -\omega ^{\prime
})+O_{ND}(\omega ,\omega ^{\prime }))\left\vert \omega \right\rangle
\left\langle \omega ^{\prime }\right\vert d\omega d\omega ^{\prime }
\label{c3}
\end{equation}%
where $O_{D}$ and $O_{ND\text{ }}$are regular functions. These observables
are contained in the space $\mathcal{O}$ of self-adjoint operators. The
introduction of distributions like $\delta (\omega -\omega ^{\prime })$ is
necessary because the \textquotedblleft singular term\textquotedblright\ $%
O_{D}(\omega )\delta (\omega -\omega ^{\prime })$ appears in observables
that cannot be left outside the space of observables, like the identity
operator, the operator whose eigenvectors are $\left\vert \omega
\right\rangle $, or the operators that commute with the latter. So, even in
this simple case the observables contain $\delta $ functions (while in more
elaborated cases they will also contain other kind of distributions).

Symmetrically, a generalized state reads:

\begin{equation}
\rho =\dint \dint (\rho _{D}(\omega )\delta (\omega -\omega ^{\prime })+\rho
_{ND}(\omega ,\omega ^{\prime }))\left\vert \omega \right\rangle
\left\langle \omega ^{\prime }\right\vert d\omega d\omega ^{\prime }
\label{c6}
\end{equation}%
where $\rho _{D}$ and $\rho _{ND}$ are regular functions. This state is
contained in a convex set of states $\mathcal{S}$. The introduction of
distributions like $\delta (\omega -\omega ^{\prime })$ is also necessary in
this case because the \textquotedblright singular term\textquotedblright\ $%
\rho _{D}(\omega )\delta (\omega -\omega ^{\prime })$ appears in generalized
states that cannot be left outside of the set $\mathcal{S}$, like the
equilibrium state.

The mean value of the observable $O$ in the state $\rho $ reads:%
\begin{gather}
Tr(\rho O)=\delta (0)\dint \rho _{D}(\omega )O_{D}(\omega )d\omega +\dint
O_{ND}(\omega ,\omega )\rho _{D}(\omega )d\omega +  \label{c9} \\
\dint \rho _{ND}(\omega ,\omega )O_{D}(\omega )d\omega +\dint \dint \rho
_{ND}(\omega ,\omega ^{\prime })O_{ND}(\omega ^{\prime },\omega )d\omega
d\omega ^{\prime }  \notag
\end{gather}%
But this result is meaningless because a term proportional to $\delta (0)$
appears.

This means that the mathematical formalism to describe continuous quantum
systems contain divergences which have no sense from the mathematical point
of view. From the just introduced mathematical formalism we can see that the
divergence can be avoided by the following transformation acting on the
state:

\begin{equation}
\Pi (\rho )=\rho -\dint \lambda (\omega )\left\vert \omega \right\rangle
\left\langle \omega \right\vert d\omega  \label{c10}
\end{equation}%
where $\lambda (\omega )$ is some regular function of $\omega $. In matrix
terms, this transformation in the discrete case acts as a displacement of
the diagonal elements:%
\begin{equation}
\left\langle u\left\vert \Pi (\rho )\right\vert v\right\rangle =\left\langle
u\left\vert \rho \right\vert v\right\rangle -\lambda (u)\delta _{uv}
\label{c10.1}
\end{equation}%
Applying again the transformation we obtain 
\begin{equation}
\Pi ^{2}(\rho )=\Pi (\Pi (\rho ))=\Pi (\rho -\dint\limits_{{}}^{{}}\lambda
(\omega )\left\vert \omega \right\rangle \left\langle \omega \right\vert
d\omega )  \label{c11}
\end{equation}%
if the transformation is linear then\footnote{%
Where linear means $\Pi (a+b)=\Pi (a)+\Pi (b)$}

\begin{equation}
\Pi (\rho -\lambda (\omega )\left\vert \omega \right\rangle \left\langle
\omega \right\vert d\omega )=\Pi (\rho )-\Pi (\dint\limits_{{}}^{{}}\lambda
(\omega )\left\vert \omega \right\rangle \left\langle \omega \right\vert
d\omega )=\Pi (\rho )  \label{c11.1}
\end{equation}%
because $\Pi (\dint\limits_{{}}^{{}}\lambda (\omega )\left\vert \omega
\right\rangle \left\langle \omega \right\vert d\omega )$ is zero, then

\begin{equation}
\Pi ^{2}(\rho )=\Pi (\rho )  \label{c11.2}
\end{equation}%
which implies that the transformation is idempotent, so it can be considered
a projector. Choosing as a regular function $\lambda (\omega )=\rho
_{D}(\omega )$, the transformation on the state reads

\begin{equation}
\Pi (\rho )=\rho -\dint \rho _{D}(\omega )\left\vert \omega \right\rangle
\left\langle \omega \right\vert d\omega =\dint \dint \rho _{ND}(\omega
,\omega ^{\prime })\left\vert \omega \right\rangle \left\langle \omega
^{\prime }\right\vert d\omega d\omega ^{\prime }  \label{c12}
\end{equation}%
Finally, the trace gives:

\begin{equation}
Tr(\Pi (\rho )O)=\dint \rho _{ND}(\omega ,\omega )O_{D}(\omega )d\omega
+\dint \dint \rho _{ND}(\omega ,\omega ^{\prime })O_{ND}(\omega ^{\prime
},\omega )d\omega d\omega ^{\prime }  \label{c13}
\end{equation}%
This is a simple example of what will be done below.

It should be clear that the divergences in the mean value of an observable
has been solved in \cite{castagnino} based in the mathematical structure
introduced in paper \cite{antoniu}. But for the purpose of this paper we
will only work with the divergences and the projector. It will be a source
of future works to describe a finite quantum field theory from the beginning
using the ideas in \cite{castagnino}.

\section{Quantum continuous systems: A general formalism for divergences}

In this section we will introduce a general formalism in terms of states and
observables following the same procedure used in decoherence. For the sake
of simplicity a few assumptions will be introduced in order to apply them to
Quantum Field Theory of a perturbative $\phi ^{4}$ theory.

The complete quantum system will be defined by $S=S_{ext}\cup S_{1}\cup
...\cup S_{p}$ where $S_{ext}$ will be called the external quantum system
and $S_{1},...,S_{p}$ will be called the internal quantum systems. The
corresponding Hilbert space is $\mathcal{H=\mathcal{H}}_{ext}\otimes 
\mathcal{\mathcal{H}}_{1}\otimes ...\otimes \mathcal{\mathcal{H}}_{p}$. Each
quantum system will contribute with diagonal and non-diagonal parts in the
observables and states in the same way as in the decoherence approach (see
section 2.2). We will make the following simplifications: we will only
consider non-diagonal observables in $S_{ext}$ and diagonal observables in
the internal quantum systems. For the states we will only consider the
non-diagonal part in the external quantum system $S_{ext}$ and both diagonal
and non-diagonal parts in the rest of the internal quantum systems. This
particular choice will be clearer below.

This means that observables and states read:

\begin{equation}
O_{rel}^{(p)}=\dint O_{ext}(x_{1},x_{2})\underset{i=1}{\overset{p}{\dprod }}%
\delta (y_{i}-w_{i})\left\vert x_{1},y_{1},...,y_{p}\right\rangle
\left\langle x_{2},w_{1},..,w_{p}\right\vert
d^{4}x_{1}d^{4}x_{2}d^{4}y_{1}d^{4}w_{1}...d^{4}y_{p}d^{4}w_{p}  \label{d1}
\end{equation}%
where the subscript $rel$ means \textquotedblleft relevant\textquotedblright 
\footnote{%
This particular name will be explained later.} and%
\begin{gather}
\rho ^{(p)}=\underset{k=0}{\overset{p-1}{\sum }}\dint \rho
_{ext}^{(k)}(x_{1},x_{2})\underset{i=1}{\overset{p}{\dprod }}\left( \rho
_{D}^{(i,k)}(y_{i})\delta (y_{i}-w_{i})+\rho
_{ND}^{(i,k)}(y_{i},w_{i})\right)  \label{d2} \\
\left\vert x_{1},y_{1},...,y_{p}\right\rangle \left\langle
x_{2},w_{1},..,w_{p}\right\vert
d^{4}x_{1}d^{4}x_{2}d^{4}y_{1}d^{4}w_{1}...d^{4}y_{p}d^{4}w_{p}  \notag
\end{gather}%
where $\left\{ \left\vert x_{1}\right\rangle \right\} $ is a continuous
basis of $\mathcal{H}_{ext}$ (and $\left\{ \left\langle x_{2}\right\vert
\right\} $ the corresponding dual basis) and each $\left\{ \left\vert
y_{p}\right\rangle \right\} $ is a basis of $\mathcal{H}_{p}$ (and $\left\{
\left\langle w_{p}\right\vert \right\} $ the corresponding dual basis). The $%
p$ superscript on the state indicates the number of internal quantum systems
and the sum in $k$ will be associated with irreducible diagrams in the
perturbation theory (this will be explained in the following sections).

The product $\rho ^{(p)}O_{rel}^{(p)}$ reads:

\begin{gather}
\rho ^{(p)}O_{rel}^{(p)}=\underset{k=0}{\overset{p-1}{\sum }}\dint \rho
_{ext}^{(k)}(x_{1},x_{2})O_{ext}((x_{2},x_{2}^{\prime })\underset{i=1}{%
\overset{p}{\dprod }}\left( \rho _{D}^{(i,k)}(y_{i})\delta
(y_{i}-w_{i})+\rho _{ND}^{(i,k)}(y_{i},w_{i})\right) \left\vert
x_{1},y_{1},...,y_{p}\right\rangle \left\langle x_{2}^{\prime
},w_{1},...,w_{p}\right\vert  \label{d3} \\
d^{4}x_{1}d^{4}x_{2}d^{4}x_{2}^{\prime
}d^{4}y_{1}...d^{4}y_{p}d^{4}w_{1}...d^{4}w_{p}  \notag
\end{gather}%
then\footnote{%
In the following equation a $\delta (0)$ appears, which is not a well
defined mathematical object. However, this fact indicates that the formalism
introduced above has a bad short-distance behaviour.}

\begin{equation}
Tr(\rho ^{(p)}O_{rel}^{(p)})=\underset{k=0}{\overset{p-1}{\sum }}\dint \rho
_{ext}^{(k)}(x_{1},x_{2})O_{ext}(x_{2},x_{1})\underset{i=1}{\overset{p}{%
\dprod }}\left( \rho _{D}^{(i,k)}(y_{i})\delta (0)+\rho
_{ND}^{(i,k)}(y_{i},y_{i})\right) d^{4}x_{1}d^{4}x_{2}d^{4}y_{1}...d^{4}y_{p}
\label{d4}
\end{equation}%
We can further on simplify the computation: in eq.(\ref{d4}) we can
calculate the integral over the $y_{i}$ coordinates as:

\begin{equation}
\dint \underset{i=1}{\overset{p}{\dprod }}\left( \rho
_{D}^{(i,k)}(y_{i})\delta (0)+\rho _{ND}^{(i,k)}(y_{i},y_{i})\right)
d^{4}y_{1}...d^{4}y_{p}=\underset{i=1}{\overset{p}{\dprod }}\dint \left(
\rho _{D}^{(i,k)}(y_{i})\delta (0)+\rho _{ND}^{(i,k)}(y_{i},y_{i})\right)
d^{4}y_{i}  \label{d4.2}
\end{equation}%
That is, the integral and the product commute, because each integrand does
not mix the coordinates. Now, we can write

\begin{equation}
\delta (0)\dint \rho _{D}^{(i,k)}(y_{i})d^{4}y_{i}+\dint \rho
_{ND}^{(i,k)}(y_{i},y_{i})d^{4}y_{i}=\delta (0)\rho _{D}^{(i,k)}+\rho
_{ND}^{(i,k)}  \label{d4.3}
\end{equation}%
where

\begin{equation}
\rho _{D}^{(i,k)}=\dint \rho _{D}^{(i,k)}(y_{i})d^{4}y_{i}\text{ \ \ \ \ \ \
\ \ \ \ \ \ \ \ \ \ }\rho _{ND}^{(i,k)}=\dint \rho
_{ND}^{(i,k)}(y_{i},y_{i})d^{4}y_{i}  \label{d4.4}
\end{equation}%
Then the r.h.s. of eq.(\ref{d4.2}) reads

\begin{equation}
\underset{i=1}{\overset{p}{\dprod }}\left( \delta (0)\rho _{D}^{(i,k)}+\rho
_{ND}^{(i,k)}\right) =\left( \delta (0)\rho _{D}^{(1,k)}+\rho
_{ND}^{(1,k)}\right) \left( \delta (0)\rho _{D}^{(2,k)}+\rho
_{ND}^{(2,k)}\right) ...\left( \delta (0)\rho _{D}^{(p,k)}+\rho
_{ND}^{(p,k)}\right)  \label{d5}
\end{equation}%
which can be written as:

\begin{equation}
\underset{i=1}{\overset{p}{\dprod }}\left( \delta (0)\rho _{D}^{(i,k)}+\rho
_{ND}^{(i,k)}\right) =\underset{l=0}{\overset{p}{\sum }}\gamma _{l}^{(p,k)}%
\left[ \delta (0)\right] ^{l}  \label{d5.1}
\end{equation}%
where

\begin{equation}
\gamma _{l}^{(p,k)}=\underset{m=1}{\overset{\binom{p}{l}}{\sum }}%
f_{m}^{(p,k,l)}  \label{d5.1.1}
\end{equation}%
where $\binom{p}{l}=\frac{p!}{l!(p-l)!}$. In particular

\begin{equation}
\ \gamma _{0}^{(p,k)}=\underset{m=1}{\overset{1}{\sum }}f_{m}^{(p,k,0)}=%
\underset{i=1}{\overset{p}{\dprod }}\rho _{ND}^{(i,k)}\text{ \ \ , .... , }\
\gamma _{p}^{(p,k)}=\underset{m=1}{\overset{1}{\sum }}f_{m}^{(p,k,p)}=%
\underset{i=1}{\overset{p}{\dprod }}\rho _{D}^{(i,k)}  \label{d5.2}
\end{equation}

All the terms $\gamma _{l}^{(p,k)}$ with $l>0$ that are multiplied by $\left[
\delta (0)\right] ^{l}$ contain at least one $\rho _{D}^{(i,k)}$, that is,
the diagonal part of the state of the $\dot{i}-$internal quantum system.

Finally, we can write:

\begin{equation}
Tr(\rho _{ext}^{(k)}O_{ext})=\dint \rho
_{ext}^{(k)}(x_{1},x_{2})O_{ext}(x_{2},x_{1})d^{4}x_{1}d^{4}x_{2}
\label{d8.1}
\end{equation}%
then eq.(\ref{d4}) reads

\begin{equation}
Tr(\rho ^{(p)}O_{rel}^{(p)})=\underset{k=0}{\overset{p-1}{\sum }}\underset{%
l=0}{\overset{p}{\sum }}\gamma _{l}^{(p,k)}\left[ \delta (0)\right]
^{l}Tr(\rho _{ext}^{(k)}O_{ext})  \label{d8.1.1}
\end{equation}

Finally, we can multiply $Tr(\rho ^{(p)}O_{rel}^{(p)})$ by $\frac{i^{p}}{p!}$
and sum over the index $p$:\footnote{%
The coefficients $\frac{i^{p}}{p!}$ are introduced for later convenience.}

\begin{equation}
Tr(\rho O_{ext})=\underset{p=0}{\overset{\infty }{\sum }}\frac{i^{p}}{p!}%
Tr(\rho ^{(p)}O_{rel}^{(p)})  \label{d8.1.2}
\end{equation}%
As we shall see in the following sections, this function $Tr(\rho O_{ext})$
is identical to the generating functional of $\phi ^{4}$ for two external
points.

Introducing eq.(\ref{d8.1.1}) in eq.(\ref{d8.1.2}) we finally have:

\begin{equation}
Tr(\rho O_{ext})=\underset{p=0}{\overset{\infty }{\sum }}\underset{k=0}{%
\overset{p-1}{\sum }}\underset{l=0}{\overset{p}{\sum }}\frac{i^{p}}{p!}%
\gamma _{l}^{(p,k)}\left[ \delta (0)\right] ^{l}Tr(\rho _{ext}^{(k)}O_{ext})
\label{d9}
\end{equation}%
This last equation can be rewritten as

\begin{equation}
Tr(\rho O_{ext})=\overset{\infty }{\underset{k=0}{\sum }}B_{k}Tr(\rho
_{ext}^{(k)}O_{ext})\text{ \ \ \ \ \ \ \ \ \ }B_{k}=\underset{l=1}{\overset{%
\infty }{\sum }}\underset{j=0}{\overset{l}{\sum }}\frac{i^{l+k}}{(l+k)!}%
\gamma _{j}^{(l,k)}\left[ \delta (0)\right] ^{j}  \label{d10}
\end{equation}%
we can obtain the last equation defining a state:

\begin{equation}
\rho =\overset{\infty }{\underset{k=0}{\sum }}B_{k}\rho ^{(k)}  \label{d11}
\end{equation}%
which resembles to a spectral decomposition of the quantum state. Finally,
we can rearrange eq.(\ref{d10})\ as

\begin{equation}
Tr(\rho O_{ext})=\underset{s=0}{\overset{\infty }{\sum }}D_{s}\left[ \delta
(0)\right] ^{s}  \label{d12}
\end{equation}%
where

\begin{equation}
D_{s}=\underset{k=0}{\overset{\infty }{\sum }}\underset{v=1}{\overset{\infty 
}{\sum }}\frac{i^{v+k}}{(v+k)!}\gamma _{s}^{(v,k)}Tr(\rho
_{ext}^{(k)}O_{ext})  \label{d13}
\end{equation}%
From this point of view, the finite contribution to the mean value of the
observable $O_{ext}$ on the state $\rho $ comes from the $s=0$ term in eq.(%
\ref{d12}) only.

\subsection{Cancellation of the divergent structure by a transformation}

We can make the following transformation in eq.(\ref{d12}):

\begin{equation}
D_{0}=\overline{D}_{0}-\underset{s=1}{\overset{\infty }{\sum }}\overline{D}%
_{s}\left[ \delta (0)\right] ^{s}  \label{d14}
\end{equation}%
then, eq.(\ref{d12}) reads:

\begin{equation}
Tr(\overline{\rho }O_{ext})=\overline{D}_{0}+\underset{s=1}{\overset{\infty }%
{\sum }}\left( D_{s}-\overline{D}_{s}\right) \left[ \delta (0)\right] ^{s}
\label{d15}
\end{equation}%
where $\overline{\rho }$ is the corresponding transformed state. If%
\begin{equation}
D_{s}-\overline{D}_{s}=0  \label{d15.1}
\end{equation}%
then%
\begin{equation}
Tr(\overline{\rho }O_{ext})=\overline{D}_{0}  \label{d15.2}
\end{equation}%
where only the finite zero order terms remains. In turn, using eq.(\ref{d13}%
) and eq.(\ref{d14}), the transformed coefficients $\overline{\gamma }%
_{s}^{(v,k)}$ of eq.(\ref{d5.2}) must obey the following equation:

\begin{equation}
\overline{\gamma }_{s}^{(v,k)}-\gamma _{s}^{(v,k)}=0\text{ \ \ for }%
s=1,...,+\infty \text{, \ \ }v=1,...,+\infty \text{, \ \ \ }k=0,...,+\infty
\label{d16}
\end{equation}%
From this point of view, the finite contribution to $Tr(\overline{\rho }%
O_{ext})$ reads:

\begin{equation}
Tr(\overline{\rho }O_{ext})=\underset{k=0}{\overset{\infty }{\sum }}\underset%
{v=1}{\overset{\infty }{\sum }}\frac{i^{v+k}}{(v+k)!}\left( \underset{i=1}{%
\overset{v}{\dprod }}\rho _{ND}^{(i,k)}\right) Tr(\rho _{ext}^{(k)}O_{ext})
\label{d18}
\end{equation}%
where the transformed state reads:

\begin{equation}
\overline{\rho }=\underset{k=0}{\overset{\infty }{\sum }}\left( \underset{v=1%
}{\overset{\infty }{\sum }}\frac{i^{v+k}}{(v+k)!}\underset{i=1}{\overset{v}{%
\dprod }}\rho _{ND}^{(i,k)}\right) \rho _{ext}^{(k)}  \label{d19}
\end{equation}

From this point of view, the cancellation of the divergent terms of the
trace (see eq.(\ref{d9})) implies a transformation of the non-diagonal and
diagonal internal quantum state (see eq.(\ref{d16})) which is a relation
between the non-diagonal and diagonal states.

This procedure is similar to the renormalization procedure in conventional
QFT by introducing counterterms in the Lagrangian. In this case, the
counterterms will be defined by new quantum states $\rho _{C.T.}$ which will
have diagonal and non-diagonal part $\overline{\rho }_{D}$ and $\overline{%
\rho }_{ND}$ and will cancel the divergences through eq.(\ref{d16}). So we
will rename the transformation introduced in this section (eq.(\ref{d14}))
as Renormaliztion.

\subsection{Cancellation of the divergence by a projection}

As we have seen in the previous subsection, we can find a transformation for
the non-diagonal functions (see eq.(\ref{d16})) so that the trace results in
a finite value. On the other hand we saw that this finite result exclusively
depends on the non-diagonal quantum state, so we can construct a projector
that projects over the non-diagonal quantum state. Following eq. (\ref{c10}%
), the projector reads

\begin{gather}
\Pi _{p}(\rho ^{(p)})=\rho ^{(p)}-\underset{k=0}{\overset{p-1}{\sum }}\dint
\rho _{ext}^{(k)}(x_{1},x_{2})\left\vert x_{1}\right\rangle \left\langle
x_{2}\right\vert d^{4}x_{1}d^{4}x_{2}(\dint \rho _{D}^{(1,k)}(y_{1})\rho
_{D}^{(2,k)}(y_{2})...\rho _{D}^{(p,k)}(y_{p})\left\vert
y_{1},...,y_{p}\right\rangle \left\langle y_{1},..,y_{p}\right\vert
d^{4}y_{1}...d^{4}y_{p}  \label{finite1} \\
+\dint \rho _{D}^{(1,k)}(y_{1},w_{1})\rho _{D}^{(2,k)}(y_{2})...\rho
_{D}^{(p-1,k)}(y_{p-1})\rho _{ND}^{(p,k)}(y_{p})\left\vert
y_{1},...,y_{p}\right\rangle \left\langle w_{1},..,y_{p}\right\vert
d^{4}y_{1}...d^{4}y_{p}d^{4}w_{1}+...  \notag \\
...+\dint \rho _{D}^{(1,k)}(y_{1},w_{1})\rho
_{ND}^{(2,k)}(y_{2},w_{2})...\rho _{ND}^{(p,k)}(y_{p},w_{p})\left\vert
y_{1},...,y_{p}\right\rangle \left\langle w_{1},..,w_{p}\right\vert
d^{4}y_{1}...d^{4}y_{p}d^{4}w_{1}...d^{4}w_{p})  \notag
\end{gather}%
This projector acting on the state $\rho ^{(p)}$ gives the following result:

\begin{equation}
\Pi _{p}(\rho ^{(p)})=\underset{k=0}{\overset{p-1}{\sum }}\dint \rho
_{ext}^{(k)}(x_{1},x_{2})\underset{i=1}{\overset{p}{\dprod }}\rho
_{ND}^{(i,k)}(y_{i},w_{i})\left\vert x_{1},y_{1},...,y_{p}\right\rangle
\left\langle x_{2},w_{1},..,w_{p}\right\vert
d^{4}x_{1}d^{4}x_{2}d^{4}y_{1}d^{4}w_{1}...d^{4}y_{p}d^{4}w_{p}
\label{finite2}
\end{equation}%
Then, the mean value of $O_{rel}^{(p)}$ in the state $\Pi _{p}(\rho ^{(p)})$
reads:

\begin{equation}
Tr(\Pi _{p}(\rho ^{(p)})O_{rel}^{(p)})=\underset{k=0}{\overset{p-1}{\sum }}%
\dint \rho _{ext}^{(k)}(x_{1},x_{2})O_{ext}(x_{2},x_{1})\underset{i=1}{%
\overset{p}{\dprod }}\rho
_{ND}^{(i,k)}(y_{i},y_{i})d^{4}x_{1}d^{4}x_{2}d^{4}y_{1}...d^{4}y_{p}
\label{finite3}
\end{equation}%
from eq.(\ref{d4.2}) and eq.(\ref{d8.1}), the last equation can be written as

\begin{equation}
Tr(\Pi _{p}(\rho ^{(p)})O_{rel}^{(p)})=\underset{k=0}{\overset{p-1}{\sum }}%
\gamma _{0}^{(p,k)}Tr(\rho _{ext}^{(k)}O_{ext})  \label{finite4}
\end{equation}%
multiplying by $\frac{i^{p}}{p!}$ and summing in $p$ we finally obtain:

\begin{equation}
Tr(\rho \Pi _{p}O_{ext})=\underset{p=0}{\overset{\infty }{\sum }}\underset{%
k=0}{\overset{p-1}{\sum }}\frac{i^{p}}{p!}\gamma _{0}^{(p,k)}Tr(\rho
_{ext}^{(k)}O_{ext})  \label{finite5}
\end{equation}%
where $\rho \Pi =\Pi (\rho ^{(p)})$ because $\Pi $ is a projector. The last
equation is similar to eq.(\ref{d10}) and we have

\begin{equation}
Tr(\rho \Pi _{p}O_{ext})=\overset{\infty }{\underset{k=0}{\sum }}B_{\Pi
_{p}}(k)Tr(\rho _{ext}^{(k)}O_{ext})\text{ \ \ \ \ \ \ \ \ \ }B_{\Pi
_{p}}(k)=\underset{l=1}{\overset{\infty }{\sum }}\frac{i^{l+k}}{(l+k)!}%
\gamma _{0}^{(l,k)}  \label{finite6}
\end{equation}%
which implies that

\begin{equation}
\rho \Pi _{p}=\overset{\infty }{\underset{k=0}{\sum }}B_{\Pi }(k)\rho
_{ext}^{(k)}  \label{finite7}
\end{equation}%
Finally, in terms of eq.(\ref{d12}), eq.(\ref{finite5}) reads

\begin{equation}
Tr(\rho \Pi _{p}O_{ext})=D_{0}  \label{finite8}
\end{equation}%
In this way, we have eliminated all the divergences of the mathematical
formalism by the application of the projector over a well defined Hilbert
subspace.\footnote{%
In section 6 we will be more precise about this Hilbert subspace.} This
formalism will be applied to the $\phi ^{4}$ theory in terms of states and
observables and then we will use dimensional regularization to localize the
divergences. Then we will show that these divergences appear in $\phi ^{4}$
with the same structure of eq.(\ref{d9}), where $[\delta (0)]^{\alpha }$
will be represented by a factor $\frac{\beta }{(d-4)^{\alpha }}$, where $d$
is the dimension of space-time.

\section{$\protect\phi ^{4}$ at first order in perturbation theory}

This section has the purpose to see how the formalism introduced in the last
section can be applied to the $\phi ^{4}$ theory at the first order in the
perturbation theory. In the Appendix B it is shown how to handle all the
other orders using dimensional regularization. We will only consider the
generating functional of two external points. In QFT, this generating
functional is simbolized $Z_{2}\left[ J\right] $, and in this case it is a
function of two external points $x_{1}$ and $x_{2}$ (see \cite{Haag},
eq.(II.2.31)):

\begin{equation}
Z_{2}\left[ J\right] =\dint \dint \tau
^{(2)}(x_{1},x_{2})J(x_{1})J(x_{2})d^{4}x_{1}d^{4}x_{2}  \label{p1}
\end{equation}%
where $\tau ^{(2)}(x_{1},x_{2})$ is the two-point connected correlation
function of the interacting theory and $J(x)$ is the source term.

The first order in the perturbation expansion of $\tau ^{(2)}(x_{1},x_{2})$
reads:

\begin{equation}
\tau ^{(2)}(x_{1},x_{2})=(-i\frac{\lambda }{4!})\dint d^{4}y_{1}\left\langle
\Omega _{0}\left\vert \phi (x_{1})\phi (x_{2})\phi ^{4}(y_{1})\right\vert
\Omega _{0}\right\rangle  \label{p2}
\end{equation}%
Introducing eq.(\ref{p2}) in eq.(\ref{p1}) the generating functional $Z_{2}%
\left[ J\right] $ reads:

\begin{equation}
Z_{2}\left[ J\right] =(-i\frac{\lambda }{4!})\dint \dint \dint \left\langle
\Omega _{0}\left\vert \phi (x_{1})\phi (x_{2})\phi ^{4}(y_{1})\right\vert
\Omega _{0}\right\rangle J(x_{1})J(x_{2})d^{4}x_{1}d^{4}x_{2}d^{4}y_{1}
\label{p3}
\end{equation}%
The only connected Feynman diagram reads:

\begin{equation}
\left\langle \Omega _{0}\left\vert \phi (x_{1})\phi (x_{2})\phi
^{4}(y_{1})\right\vert \Omega _{0}\right\rangle =\Delta (x_{1}-y_{1})\Delta
(x_{2}-y_{1})\Delta (y_{1}-y_{1})  \label{p4}
\end{equation}%
where $\Delta (x-y)$ is the scalar propagator. This propagator diverges when 
$x=y$, which means that $\tau ^{(2)}$ diverges due to the factor $\Delta
(y_{1}-y_{1})$ in eq.(\ref{p4}). To formally avoid this divergence, without
changing the theory, we can introduce a Dirac delta in eq.(\ref{p4}) so

\begin{equation}
\left\langle \Omega _{0}\left\vert \phi (x_{1})\phi (x_{2})\phi
^{4}(y_{1})\right\vert \Omega _{0}\right\rangle =\dint d^{4}w_{1}\Delta
(x_{1}-y_{1})\Delta (x_{2}-y_{1})\Delta (y_{1}-w_{1})\delta (y_{1}-w_{1})
\label{p5}
\end{equation}%
Introducing eq.(\ref{p5}) in eq.(\ref{p3}) we have:

\begin{equation}
Z_{2}\left[ J\right] =12\cdot (-i\frac{\lambda }{4!})\dint \dint \dint \dint
\Delta (x_{1}-y_{1})\Delta (x_{2}-y_{1})\Delta (y_{1}-w_{1})\delta
(y_{1}-w_{1})J(x_{1})J(x_{2})d^{4}x_{1}d^{4}x_{2}d^{4}y_{1}d^{4}w_{1}
\label{p6}
\end{equation}%
where $12$ is the symmetry factor.\footnote{%
From eq.(\ref{p6}) we can interpret the quantum state as the tree diagram
associated to the Feynmann diagram of the first order in the perturbation
expansion. The propagator $\Delta (y_{1}-w_{1})$ is transformed into a loop
when we introduce the observable which has a $\delta (y_{1}-w_{1})$. This
procedure can be done for all the Feynmann diagrams, but it only introduces
a pictorical way to understand the quantum states.} We can call

\begin{equation}
\rho (x_{1},y_{1},x_{2},w_{1})=\Delta (x_{1}-y_{1})\Delta
(x_{2}-y_{1})\Delta (y_{1}-w_{1})  \label{p7}
\end{equation}%
and%
\begin{equation}
O_{ext}^{ND}(x_{1},x_{2})=J(x_{1})J(x_{2})  \label{p8}
\end{equation}%
then eq.(\ref{p6}) reads:

\begin{equation}
Z_{2}\left[ O_{ext}^{ND}\right] =12(-i\frac{\lambda }{4!})\dint \dint \dint
\dint \rho (x_{1},y_{1},x_{2},w_{1})\delta
(y_{1}-w_{1})O_{ext}^{ND}(x_{1},x_{2})d^{4}x_{1}d^{4}x_{2}d^{4}y_{1}d^{4}w_{1}
\label{p9}
\end{equation}%
which is identical to eq.(\ref{c1}) with $\omega _{S}=x_{1}$, $\omega
_{E}=y_{1}$, $\omega _{S}^{\prime }=x_{2}$ and $\omega _{E}^{\prime }=w_{1}$.

Following the notation of eq.(\ref{d4}), we can write eq.(\ref{p9}) as

\begin{equation}
Z_{2}=Tr(\rho ^{(1)}O_{rel}^{(1)})  \label{p10.1}
\end{equation}%
where

\begin{equation}
\rho ^{(1)}=\dint \dint \dint \dint \Delta (x_{1}-y_{1})\Delta
(x_{2}-y_{1})\Delta (y_{1}-w_{1})\left\vert x_{1},y_{1}\right\rangle
\left\langle x_{2},w_{1}\right\vert d^{4}x_{1}d^{4}x_{2}d^{4}y_{1}d^{4}w_{1}
\label{p10.2}
\end{equation}%
and

\begin{equation}
O_{rel}^{(1)}=\dint \dint \dint \dint J(x_{1})J(x_{2})\delta
(y_{1}-w_{1})\left\vert x_{1},y_{1}\right\rangle \left\langle
x_{2},w_{1}\right\vert d^{4}x_{1}d^{4}x_{2}d^{4}y_{1}d^{4}w_{1}
\label{p10.3}
\end{equation}%
In principle we must admit that the definition of state given by eq.(\ref%
{p10.2}) is not rigorous because $Tr(\rho ^{(1)})=\infty $. But this is
exactly what we are trying to found in the mathematical formalism of QFT.
When this problem will be resolved, we will obtain the normalization of the
state without difficulties.

\subsection{Reduced state}

As we have seen in the previous section (see eq. (\ref{d1})), the relevant
observable of eq.(\ref{p10.3}) can be written as

\begin{equation}
O_{rel}^{(1)}=O_{ext}^{ND}\otimes I_{int}  \label{p10.0}
\end{equation}%
This is analogous to the observable of eq.(\ref{EIDgen-02}). In the
continuous case, eq.(\ref{p9}) can be written as the trace of an observable
in a reduced state, analogously to eq.(\ref{c2}). To be more precise, it is
convenient to remember which the Hilbert spaces are. The external system $%
S_{ext}~$corresponds to the coordinate $x_{1}$ and $x_{2}$ and the internal
quantum system $S_{int}$ corresponds to the $y_{1}$ and $w_{1}$ coordinates.
The composite system is $S=S_{ext}\cup S_{int}$ with the corresponding
Hilbert space $\mathcal{H=H}_{ext}\otimes \mathcal{H}_{int}$. The continuous
basis for $\mathcal{H}_{ext}~$is $\left\{ \left\vert x_{1}\right\rangle
\right\} $ (and the corresponding basis of the dual space is $\left\{
\left\langle x_{2}\right\vert \right\} $), and the continuous basis for $%
\mathcal{H}_{int}$ is $\left\{ \left\vert y_{1}\right\rangle \right\} $ (and
the corresponding basis of the dual space is $\left\{ \left\langle
w_{1}\right\vert \right\} $) which means that in Section 3 it is $p=1$, so $%
p $ counts the order in perturbation, the number of internal quantum systems
and the internal coordinates. Both external and internal coordinates come in
pairs. This means that the only contributions to the generating functional
comes form an even number of external and internal coordinates. This agrees
with $\phi ^{4}$ theory because the generating functional vanishes for an
odd number of external coordinates.

Then, the eq.(\ref{p10.1}) can be written as the trace of an observable in a
reduced state:%
\begin{equation}
Z_{2}\left[ O_{ext}\right] =12(-i\frac{\lambda }{4!})\dint \dint
Tr_{int}(\rho ^{(1)})O_{ext}^{ND}(x_{1},x_{2})d^{4}x_{1}d^{4}x_{2}=Tr(%
\overline{\rho }_{ext}^{(1,0)}O_{ext})  \label{p10}
\end{equation}%
where the reduced state $\overline{\rho }_{ext}^{(1,0)}$ reads\footnote{%
The bar above the state $\overline{\rho }$ indicates that this state is not
the same as the one in eq. (\ref{d9}).}:%
\begin{equation}
\overline{\rho }_{ext}^{(0)}=Tr_{int}(\rho ^{(1)})=\dint \left\langle
y_{1}^{\prime }\left\vert \rho ^{(1)}\right\vert y_{1}^{\prime
}\right\rangle d^{4}y_{1}^{\prime }=\left( \dint \Delta (x_{1}-y_{1})\Delta
(x_{2}-y_{1})\Delta (0)dy_{1}\right) \left\vert x_{1}\right\rangle
\left\langle x_{2}\right\vert d^{4}x_{1}d^{4}x_{2}  \label{p11}
\end{equation}%
where $Tr_{int}$ is the partial trace of $\rho ^{(1)}$ with respect to the
internal coordinates $y_{1}$ and $O_{ext}$ reads:\footnote{%
In eq.(\ref{p11.1})\ the source terms $J(x_{1})$ and $J(x_{2})$ acquire an
important rol in the formalism introduced above: they are the distribution
function of the external observables.}

\begin{equation}
O_{ext}=\dint J(x_{1})J(x_{2})\left\vert x_{1}\right\rangle \left\langle
x_{2}\right\vert d^{4}x_{1}d^{4}x_{2}  \label{p11.1}
\end{equation}

The reduced state of eq.(\ref{p11}) is divergent because the component of $%
\overline{\rho }_{ext}^{(0)}$ contains a $\Delta (0)$. This state must be
regularized, which means that we must extract the singular term. It is
important to note that the reduced state has a divergence because we have
taken the partial trace over the internal coordinates. This does not mean
that the reduced state, which depends on the external coordinates $x_{1}$
and $x_{2}$ is singular. In fact, because $x_{1}$ and $x_{2}$ are the
external points, they must be different $x_{1}\neq x_{2}$. So, the
divergence only comes from taking $y_{1}=w_{1}$ which is identical to have a
diagonal state in the internal quantum system, which means that in fact our
state is identical to the state of eq.(\ref{d2}) with $p=1$. This is similar
to take the internal partial trace on eq. (\ref{d2}), which gives

\begin{gather}
Tr_{int}(\rho ^{(1)})=\dint \left\langle y_{1}^{\prime }\left\vert \rho
^{(1)}\right\vert y_{1}^{\prime }\right\rangle =\delta (0)\gamma
_{1}^{(1,0)}\dint \rho _{ext}^{(1.0)}(x_{1},x_{2})\left\vert
x_{1}\right\rangle \left\langle x_{2}\right\vert d^{4}x_{1}d^{4}x_{2}+
\label{p11.2} \\
\gamma _{0}^{(1,0)}\dint \rho _{ext}^{(1,0)}(x_{1},x_{2})\left\vert
x_{1}\right\rangle \left\langle x_{2}\right\vert d^{4}x_{1}d^{4}x_{2}  \notag
\end{gather}%
where (see eq.(\ref{d4.3}) and eq.(\ref{d5.2})):

\begin{equation}
\gamma _{1}^{(1,0)}=\left( \dint \rho _{D}^{(1,0)}(y_{1})d^{4}y_{1}\right) 
\text{ \ \ \ \ \ \ \ \ \ \ \ \ \ \ \ \ \ }\gamma _{0}^{(1,0)}=\left( \dint
\rho _{ND}^{(1,0)}(y_{1},y_{1})d^{4}y_{1}\right)  \label{p11.2.1}
\end{equation}%
\ To give eq.(\ref{p10}) the form of eq.(\ref{d9}), we must regularize $%
\Delta (0)$ through dimensional regularization.

The $\Delta (\xi )$ reads (\cite{PS}, pag. 31, eq.(2.59)):

\begin{equation}
\Delta (\xi )=\dint \frac{d^{4}p}{(2\pi )^{4}}\frac{ie^{-ip\xi }}{%
p^{2}-m_{0}^{2}+i\epsilon }  \label{p12}
\end{equation}%
Then, the component of the reduced state $\overline{\rho }_{ext}^{(1,0)}$
(see eq.(\ref{p11})) reads

\begin{equation}
\overline{\rho }_{ext}^{(0)}(x_{1},x_{2})=i^{3}\dint \frac{d^{4}p}{(2\pi
)^{4}}\frac{e^{-ip(x_{1}-x_{2})}}{(p^{2}-m^{2}+i\epsilon )^{2}}\dint \frac{%
d^{4}l}{(2\pi )^{4}}\frac{1}{l^{2}-m^{2}}  \label{p13}
\end{equation}%
The $l$-momentum integral diverges when $l\rightarrow \infty $. The
dimensional regularization \cite{TW} consists to compute the Feynman diagram
as an analytical function of the dimensionality of space-time, $d$. In this
way, the $p-$ momentum integral reads 
\begin{equation}
\dint \frac{d^{d}l}{(2\pi )^{d}}\frac{1}{l^{2}-m_{0}^{2}}=\frac{m_{0}^{2}}{%
(4\pi )^{2}}\left( \frac{m_{0}^{2}}{4\pi }\right) ^{\frac{d}{2}-2}\Gamma (1-%
\frac{d}{2})  \label{p14}
\end{equation}%
where $\Gamma (1-\frac{d}{2})$ is the Gamma function which diverges when $%
d=4,6,8,...$. Near $d=4$, $\Gamma (1-\frac{d}{2})$ behaves as%
\begin{equation}
\Gamma (1-\frac{d}{2})\approx \frac{2}{\epsilon }+\gamma +O(\epsilon )
\label{p15}
\end{equation}%
where $\gamma =\frac{\pi ^{2}}{12}$ is the Euler-Mascheroni constant and $%
O(\epsilon )$ is a sum of powers in $\epsilon =d-4$.

Expanding in Taylor series the $\left( \mu ^{2}\right) ^{4-d}$ term in eq.(%
\ref{p14}) and using eq.(\ref{p15}) we have:\footnote{$\mu $ is constant
mass factor introduced to have a dimensionless coupling constant.}%
\begin{equation}
\left( \mu ^{2}\right) ^{-\epsilon }\dint \frac{d^{d}l}{(2\pi )^{d}}\frac{1}{%
l^{2}-m_{0}^{2}}=\frac{m_{0}^{2}}{(4\pi )^{2}}\left[ 1-\epsilon \ln (\frac{%
4\pi \mu }{m_{0}^{2}}^{2})+O(\epsilon )\right] \left[ \frac{2}{\epsilon }%
+\Psi (2)+O(\epsilon )\right]  \label{p16}
\end{equation}%
where $\Psi (2)=1-\gamma $, so%
\begin{equation}
\left( \mu ^{2}\right) ^{-\epsilon }\dint \frac{d^{d}l}{(2\pi )^{d}}\frac{1}{%
l^{2}-m_{0}^{2}}=\frac{m_{0}^{2}}{(4\pi )^{2}}\left[ \Psi (2)-2\ln (\frac{%
m_{0}^{2}}{4\pi \mu ^{2}})+\frac{2}{\epsilon }+O(\epsilon )\right]
\label{p17}
\end{equation}%
then the reduced state can be written as:

\begin{equation}
\overline{\rho }_{ext}^{(0)}(x_{1},x_{2})=\beta _{1}^{(1,0)}\frac{1}{%
\epsilon }\rho _{ext}^{(0)}+\beta _{0}^{(1,0)}\rho _{ext}^{(0)}  \label{p18}
\end{equation}%
where $\beta _{1}^{(1,0)}=\frac{m_{0}^{2}}{(4\pi )^{2}}$ and $\beta
_{0}^{(1,0)}=\frac{m_{0}^{2}}{(4\pi )^{2}}\ln (\frac{m_{0}^{2}}{4\pi }%
)+\gamma $ and

\begin{equation}
\rho _{ext}^{(0)}=i^{3}\dint \frac{d^{4}p}{(2\pi )^{4}}\frac{%
e^{-ip(x_{1}-x_{2})}}{(p^{2}-m_{0}^{2})^{2}}  \label{p21.1}
\end{equation}%
In the other side, if we take $p=1$ in eq.(\ref{d8.1.1})\ we obtain

\begin{equation}
Tr(\rho ^{(1)}O_{rel}^{(1)})=\underset{k=0}{\overset{0}{\sum }}\underset{l=0}%
{\overset{1}{\sum }}\gamma _{l}^{(p,k)}\left[ \delta (0)\right] ^{l}Tr(\rho
_{ext}^{(k)}O_{ext})=\left( \gamma _{0}^{(1,0)}+\gamma _{1}^{(1,0)}\delta
(0)\right) Tr(\rho _{ext}^{(0)}O_{ext})  \label{p21.2}
\end{equation}%
and we can make the following replacement

\begin{gather}
\gamma _{0}^{(1,0)}=\beta _{1}^{(1,0)}=\ \rho _{ND}^{(0)}=\frac{m_{0}^{2}}{%
(4\pi )^{2}}\text{ \ \ ,\ \ \ \ \ }\gamma _{1}^{(1,0)}=\beta
_{0}^{(1,0)}=\rho _{D}^{(0)}=\frac{m_{0}^{2}}{(4\pi )^{2}}\ln (\frac{%
m_{0}^{2}}{4\pi })+\gamma  \label{p21.3} \\
Tr(\rho
_{ext}^{(0)}O_{ext})=i^{3}\dint%
\limits_{{}}^{{}}d^{4}x_{1}d^{4}x_{2}J(x_{1})J(x_{2})\dint \frac{d^{4}p}{%
(2\pi )^{4}}\frac{e^{-ip(x_{1}-x_{2})}}{(p^{2}-m_{0}^{2})^{2}}\text{ \ \ \ }
\notag \\
\text{\ \ \ \ \ \ \ \ }R\left[ \delta (0)\right] =\underset{\epsilon
\rightarrow 0}{\lim }\frac{1}{\epsilon }  \notag
\end{gather}%
These last two equations (\ref{p21.2} and \ref{p21.3}) explain us how the
mathematical formalism introduced in Section 3 is related with the QFT of $%
\phi ^{4}$ theory. In the following section we will show how to find all the
orders in the perturbation theory.\footnote{%
It will be the subject of further work to determine the diagonal and
non-diagonal functions without making use of QFT in its original version.
This functions depends on what happens in short and long-distances. In the
first case we think that a more fundamental theory can give us the desired
result.}

The reduced state computed in eq.(\ref{p11}) has a physical counterpart. It
is well known that the reduction of a state decreases the information
available to the observer about the composite system. In the case above, the
reduction is done over the internal vertices where the interaction ocurrs.
In QFT, the particles that are created in this vertices are virtual
particles because they are off-shell, that is, they do not obey the
conservation laws. In this sense, the conceptual meaning of the partial
trace of the internal degrees of freedom is to neglect the virtual
non-physical particles. This is consistent with the experiments of
scattering because basically what is seen are the in and out states.
However, perturbation theory introduces off-shell intermediate states whose
existence depends on the uncertainty principle $\Delta E\Delta t\geq \frac{%
\hbar }{2}$. In turn, the interpretation of the integration of the internal
vertices is to sum over all points where this process can ocurr (see \cite%
{PS}, page 94). In our case, the integration over the internal vertices
reflects the fact that we are neglecting the degrees of freedom of this
virtual particles and what we finally obtain is a reduced state which is
divergent.

\subsection{The projection at first order}

To see how the projector acts at first order in perturbation theory, we can
use eq.(\ref{finite1}) with $p=1$:

\begin{equation}
\Pi _{1}(\rho ^{(1)})=\rho ^{(1)}-\dint \rho _{ext}^{(0)}(x_{1},x_{2})\rho
_{D}^{(0)}(y_{1})\left\vert x_{1},y_{1}\right\rangle \left\langle
x_{2},y_{1}\right\vert d^{4}x_{1}d^{4}x_{2}d^{4}y_{1}  \label{proo1}
\end{equation}%
where

\begin{equation}
\rho _{ext}^{(0)}(x_{1},x_{2})=\dint \frac{d^{4}p}{(2\pi )^{4}}\frac{%
e^{-ip(x_{1}-x_{2})}}{(p^{2}-m_{0}^{2})^{2}}\text{ \ \ \ \ \ \ \ \ \ \ }%
\dint\limits_{{}}^{{}}\rho _{D}^{(0)}(y_{1})d^{4}y_{1}=\beta _{0}^{(1,0)}=%
\frac{m_{0}^{2}}{(4\pi )^{2}}\ln (\frac{m_{0}^{2}}{4\pi })+\gamma
\label{proo2}
\end{equation}%
then

\begin{equation}
Tr(\rho ^{(1)}\Pi _{1}O_{rel}^{(1)})=\rho _{ND}^{(0)}Tr(\rho
_{ext}^{(0)}O_{ext})  \label{proo3}
\end{equation}%
where

\begin{equation}
\rho _{ND}^{(0)}=\frac{m_{0}^{2}}{(4\pi )^{2}}  \label{proo4}
\end{equation}%
In this way, using the formalism introduced in Section 3, we can neglect the
divergence that appears at first order in the perturbation expansion by
projecting over the finite contribution instead of introducing counterterms
in the Lagrangian.

\section{General procedure for $\protect\phi ^{4}$}

In Appendix B we will show the full $\phi ^{4}$ perturbation theory for the
two-point correlation function. For simplicity we just remember the main
result (see eq.(\ref{q6})):

\begin{equation}
\dint \left\langle \Omega \left\vert \phi (x_{1})\phi (x_{2})\right\vert
\Omega \right\rangle J(x_{1})J(x_{2})d^{4}x_{1}d^{4}x_{2}=\overset{+\infty }{%
\underset{s=0}{\sum }}\dint \frac{d^{4}p}{(2\pi )^{4}}\frac{%
e^{-ip(x_{1}-x_{2})}}{(p^{2}-m_{0}^{2})^{1+s}}%
J(x_{1})J(x_{2})d^{4}x_{1}d^{4}x_{2}\overset{+\infty }{\underset{n=0}{\sum }}%
\overset{+\infty }{\underset{j=1}{\sum }}(\frac{i\lambda }{4!})^{j}\beta
_{n}^{(j,s)}\frac{1}{\epsilon ^{n}}  \label{gen1}
\end{equation}%
If we make the following replacement in eq.(\ref{d10})

\begin{gather}
\text{(a) \ \ \ }\rho _{ext}^{(k)}=\dint \frac{d^{4}p}{(2\pi )^{4}}\frac{%
e^{-ip(x_{1}-x_{2})}}{(p^{2}-m_{0}^{2})^{1+k}}\text{ \ \ \ \ \ \ \ \ \ \ \ \
\ \ (b) \ \ \ }O_{ext}=J(x_{1})J(x_{2})  \label{hola} \\
\text{ (c) \ \ \ \ }(\frac{i\lambda }{4!})^{j}\beta _{n}^{(j,s)}=\frac{%
i^{j+s}}{(j+s)!}\gamma _{n}^{(j,s)}\text{ \ \ \ \ \ \ \ \ \ \ \ \ \ (d) \ \
\ }R(\left[ \delta (0)\right] ^{n})=\underset{\epsilon \rightarrow 0}{\lim }%
\frac{1}{\epsilon ^{n}}  \notag
\end{gather}%
we obtain

\begin{equation}
\dint \left\langle \Omega \left\vert \phi (x_{1})\phi (x_{2})\right\vert
\Omega \right\rangle J(x_{1})J(x_{2})d^{4}x_{1}d^{4}x_{2}=Tr(\rho O_{ext})
\label{gen3}
\end{equation}%
Eq.(\ref{hola}.c) gives the relation between the mathematical formalism
introduced in Section 3 and the conventional QFT\ using dimensional
regularization.

For simplicity, we will develop the following results directly using eq.(\ref%
{d10}) where

\begin{equation}
\rho _{ext}^{(k)}=\dint \frac{d^{4}p}{(2\pi )^{4}}\frac{e^{-ip(x_{1}-x_{2})}%
}{(p^{2}-m_{0}^{2})^{1+k}}  \label{gen4}
\end{equation}%
then%
\begin{equation}
Tr(\rho _{ext}^{(k)}O_{ext})=\dint \frac{d^{4}q}{(2\pi )^{4}}\frac{f(q)}{%
(q^{2}-m_{0}^{2})^{1+k}}  \label{gen4.1}
\end{equation}%
where

\begin{equation}
f(q)=\dint d^{4}x_{1}d^{4}x_{2}e^{-iq(x_{1}-x_{2})}J(x_{1})J(x_{2})
\label{gen4.2}
\end{equation}%
Introducing eq.(\ref{gen4.1}) in eq.(\ref{d10}) we have

\begin{equation}
Tr(\rho O_{ext})=\dint \frac{d^{4}q}{(2\pi )^{4}}f(q)\left( \frac{1}{%
q^{2}-m_{0}^{2}}+\underset{n=0}{\overset{\infty }{\sum }}\frac{1}{%
(q^{2}-m_{0}^{2})^{2+n}}\underset{r=1}{\overset{\infty }{\sum }}\underset{l=0%
}{\overset{r+n}{\sum }}\frac{i^{r+n}}{(r+n)!}\gamma _{l}^{(r+n,n)}\left[
\delta (0)\right] ^{l}\right) \text{\ }  \label{gen5}
\end{equation}%
if we apply the projection of section 3.2, eq. (\ref{finite1}) order by
order, we must only keep the term with $l=0$ in eq.(\ref{gen5}), then

\begin{equation}
Tr(\rho \Pi O_{ext})=\dint \frac{d^{4}q}{(2\pi )^{4}}f(q)\left( \frac{1}{%
q^{2}-m_{0}^{2}}+\underset{n=0}{\overset{\infty }{\sum }}\frac{1}{%
(q^{2}-m_{0}^{2})^{2+n}}\underset{r=1}{\overset{\infty }{\sum }}\frac{i^{r+n}%
}{(r+n)!}\gamma _{0}^{(r+n,n)}\right)  \label{gen5.1}
\end{equation}%
The first term of the r.h.s of the last equation is the propagator of the
non-interacting theory. The second term with $n=0$ contains the sum of all
one-particle irreducible diagrams $\Sigma (p)$ (see \cite{PS}, page 228,
eq.(7.43)) :

\begin{equation}
\Sigma (p)=M^{2}(0)=\underset{r=1}{\overset{\infty }{\sum }}\frac{i^{r}}{(r)!%
}\gamma _{0}^{(r,0)}  \label{gen5.2}
\end{equation}%
In fact, the following terms with $n>1$ in eq.(\ref{gen5.1}) are the product
of one-particle irreducible diagrams $\Sigma (p)$, which means that

\begin{equation}
M^{2}(n)=\left[ M^{2}(0)\right] ^{n+1}  \label{gen5.3}
\end{equation}%
this gives a relation between the coefficients $\gamma _{n}^{(r+n,n)}$ and $%
\gamma _{0}^{(r,0)}$:

\begin{equation}
\underset{r=1}{\overset{+\infty }{\sum }}\frac{i^{r+n}}{(r+n)!}\gamma
_{0}^{(r+n,n)}=\left( \underset{r=1}{\overset{+\infty }{\sum }}\frac{i^{r}}{%
r!}\gamma _{0}^{(r,0)}\right) ^{n+1}  \label{gen5.4}
\end{equation}%
For example, for $n=1$, eq.(\ref{gen5.4}) implies that:

\begin{equation}
\underset{j=1}{\overset{r}{\sum }}\frac{\gamma _{0}^{(j,0)}\gamma
_{0}^{(r-j+1,0)}}{j!(r-j+1)!}=\frac{\gamma _{0}^{(r+1,1)}}{(r+1)!}
\label{gen5.5}
\end{equation}%
from eq.(\ref{gen5.3}) and eq.(\ref{gen5.1}) we have

\begin{equation}
Tr(\rho \Pi O_{ext})=\dint \frac{d^{4}q}{(2\pi )^{4}}f(q)\left( \frac{1}{%
q^{2}-m_{0}^{2}}\underset{n=0}{\overset{\infty }{\sum }}\left( \frac{M^{2}(0)%
}{q^{2}-m_{0}^{2}}\right) ^{n}\right)  \label{gen5.6}
\end{equation}%
if $\left\vert \frac{M^{2}(0)}{q^{2}-m_{0}^{2}}\right\vert <1$ then

\begin{equation}
\underset{n=0}{\overset{\infty }{\sum }}\left( \frac{M^{2}(0)}{%
q^{2}-m_{0}^{2}}\right) ^{n}=\frac{1}{1-\frac{M^{2}(0)}{q^{2}-m_{0}^{2}}}
\label{gen5.7}
\end{equation}%
introducing eq.(\ref{gen5.7}) in eq.(\ref{gen5.6})\ we finally obtain

\begin{equation}
Tr(\rho \Pi O_{ext})=\dint \frac{d^{4}q}{(2\pi )^{4}}\frac{f(q)}{%
q^{2}-(m_{0}^{2}+M^{2}(0))}  \label{gen5.8}
\end{equation}%
where the pole in the mass value has been shifted away by an amount of $%
M^{2}(0)$. If we do keep all the $l\,\ $terms in eq.(\ref{gen5.1}) we have

\begin{equation}
Tr(\rho O_{ext})=\dint \frac{d^{4}q}{(2\pi )^{4}}f(q)\left( \frac{1}{%
q^{2}-m_{0}^{2}}+\underset{n=0}{\overset{\infty }{\sum }}\frac{1}{%
(q^{2}-m_{0}^{2})^{2+n}}L^{2}(n)\right)  \label{gen5.9}
\end{equation}%
where

\begin{equation}
L^{2}(n)=\underset{r=1}{\overset{\infty }{\sum }}\underset{l=0}{\overset{r+n}%
{\sum }}\frac{i^{r+n}}{(r+n)!}\gamma _{l}^{(r+n,n)}\left[ \delta (0)\right]
^{l}  \label{gen5.10}
\end{equation}%
then if we introduce the condition

\begin{equation}
L^{2}(n)=\left[ L^{2}(0)\right] ^{n+1}  \label{gen5.11}
\end{equation}%
we have

\begin{equation}
Tr(\rho O_{ext})=\dint \frac{d^{4}q}{(2\pi )^{4}}f(q)\underset{n=0}{\overset{%
\infty }{\sum }}\frac{\left[ L^{2}(0)\right] ^{n}}{(q^{2}-m_{0}^{2})^{1+n}}%
=\dint \frac{d^{4}q}{(2\pi )^{4}}\frac{f(q)}{q^{2}-(m_{0}^{2}+L^{2}(0))}
\label{gen5.12}
\end{equation}%
where we have written that $\left\vert \frac{L^{2}(0)}{q^{2}-m_{0}^{2}}%
\right\vert <1$ which, of course, has no sense because $L^{2}(0)$ is
divergent (see eq.(\ref{gen5.10})). Nevertheless, the mass shift reads

\begin{equation}
m^{2}=m_{0}^{2}+\underset{r=1}{\overset{\infty }{\sum }}\frac{i^{r}}{r!}%
\gamma _{0}^{(r,0)}+\underset{r=1}{\overset{\infty }{\sum }}\underset{l=1}{%
\overset{r}{\sum }}\frac{i^{r}}{r!}\gamma _{l}^{(r,0)}\left[ \delta (0)%
\right] ^{l}  \label{gen5.13}
\end{equation}%
which is identical to eq.(\ref{q16}) and to eq.(2.3a) of \cite{thooft}.

Given the relation of eq.(\ref{gen5.13}) and eq.(\ref{q16}), the
renormalization group is hidden in the last equation, because we have not
introduced some constants like the mass factor, which is inside the
functions $\beta _{n}^{(j,0)}$ and $\beta _{0}^{(j,0)}$ in Appendix B.

\section{The projection in algebraic terms}

We can rewrite Section 3 in algebraic languaje, then, for each order in the
perturbation theory we have the following Hilbert spaces:

\begin{eqnarray}
p &=&0\text{ \ \ \ \ \ \ \ \ \ \ }\mathcal{H}^{(0)}=\mathcal{H}_{ext}\text{
\ \ \ \ \ \ \ \ \ \ \ \ \ \ \ \ \ \ \ \ \ \ \ \ \ \ \ \ \ \ }  \label{sys1}
\\
p &=&1\text{ \ \ \ \ \ \ \ \ \ \ }\mathcal{H}^{(1)}=\mathcal{H}_{ext}\otimes 
\mathcal{H}_{int}^{(1)}  \notag \\
&&\vdots \text{ \ \ \ \ \ \ \ \ \ \ \ \ \ }\vdots  \notag \\
p &=&j\text{ \ \ \ \ \ \ \ \ \ \ \ }\mathcal{H}^{(j)}=\mathcal{H}%
_{ext}\otimes \mathcal{H}_{int}^{(1)}\otimes ...\otimes \mathcal{H}%
_{int}^{(j)}\text{\ }  \notag
\end{eqnarray}%
The total Hilbert space to all orders in the perturbation theory reads

\begin{equation}
\mathcal{H}=\mathcal{H}^{(0)}\oplus \mathcal{H}^{(1)}\oplus ...\oplus 
\mathcal{H}^{(p)}=\underset{i=0}{\overset{p}{\oplus }}\mathcal{H}^{(i)}
\label{sys2}
\end{equation}

The algebra of observables $\mathcal{O}$ is represented by $^{\ast }-$%
algebra $\mathcal{A}$ of self-adjoint elements and states are represented by
functionals on $\mathcal{O}$, that is, by elements of the dual space $%
\mathcal{O}^{\prime }$, $\rho \in \mathcal{O}^{\prime }$. In this work, we
will shall adopt a $C^{\ast }-$algebra of operators. As it is well known, a $%
C^{\ast }-$algebra can be represented in a Hilbert space $\mathcal{H}$ (GNS\
theorem) and, in this particular case $\mathcal{O=\mathcal{O}}^{\prime }$;
therefore$\mathcal{\ \mathcal{O}}$ and $\mathcal{O}^{\prime }$ are
represented by $\mathcal{H\otimes \mathcal{H}}$ that will be called $%
\mathcal{N}$ which reads

\begin{equation}
\mathcal{N=H\otimes H=}(\mathcal{H}^{(0)}\otimes \mathcal{H}^{(0)})\oplus
...\oplus (\mathcal{H}^{(p)}\otimes \mathcal{H}^{(p)})=\mathcal{N}%
^{(0)}\oplus ...\oplus \mathcal{N}^{(p)}  \label{sys3}
\end{equation}%
Now let $\mathcal{N}_{S}$ be the space of singular parts (namely the one
containing the $\delta (x)$) and $\mathcal{N}_{R}$ the space of the regular
parts (namely the non-diagonal part\textbf{)} of $\mathcal{\mathcal{\mathcal{%
N}}}$.

Then 
\begin{equation}
\mathcal{N}_{S},\mathcal{N}_{R}\subset \mathcal{\mathcal{\mathcal{N}}}
\label{3.14}
\end{equation}%
We can make the quotient 
\begin{equation}
\frac{\mathcal{\mathcal{\mathcal{N}}}}{\mathcal{N}_{S}}=\mathcal{N}_{R}
\label{i}
\end{equation}%
where $\mathcal{N}_{R}$ would represent the vector space of equivalent
classes of non-diagonal operators. These equivalence classes read 
\begin{equation}
\lbrack a]=a+\mathcal{N}_{S},\qquad a\in \mathcal{\mathcal{\mathcal{N}}}
\label{j}
\end{equation}%
So we can decompose $\mathcal{\mathcal{\mathcal{N}}}$ as: 
\begin{equation}
\mathcal{\mathcal{\mathcal{N}}=N}s+\mathcal{N}_{R}  \label{A3}
\end{equation}%
But eq. (\ref{j}) is not a direct sum, since we can add an arbitrary $a\in 
\mathcal{N}_{S}$ from the first term of the r. h. s. of the last equation
and substract $a$ from the second term.

As we are interested in the diagonal and non-diagonal elements of a matrix
state we can define a sub algebra of $\mathcal{\mathcal{\mathcal{N}}}$, that
can be called a van Hove algebra \cite{van hove} since it is inspired in the
works of this author, as:

\begin{equation}
\mathcal{\mathcal{\mathcal{N}}}_{vh}\mathcal{=N}_{S}\oplus \mathcal{N}%
_{R}\subset \mathcal{\mathcal{\mathcal{N}}}  \label{k}
\end{equation}%
where the vector space $\mathcal{N}_{R}$ is the space of operators with $%
O(x)=0$ and $O(x,x^{\prime })$ is a regular function. Moreover $\mathcal{O=N}%
_{vhS}$ is the space of selfadjoint operators of $\mathcal{\mathcal{\mathcal{%
N}}}_{vh},$ which can be constructed in such a way it could be dense in $%
\mathcal{N}_{S}$ (because any distribution can be approximated by regular
functions). Therefore essentially the introduced restriction is the minimal
possible coarse-graining. Now the $\oplus $ is a direct sum because $%
\mathcal{N}_{S}$ contains the factor $\delta (x-x^{\prime })$ and $\mathcal{N%
}_{R}$ contains just regular functions and a kernel cannot be both a $\delta 
$ and a regular function. Moreover, as our observables must be self-adjoint,
the space of observables must be 
\begin{equation}
\mathcal{O=N}_{vhS}\mathcal{=N}_{S}\oplus \mathcal{N}_{R}\subset \mathcal{N}
\label{l}
\end{equation}%
This decomposition corresponds to the one given in eq. (\ref{c3}) or eq.(\ref%
{d1}) where $\mathcal{N}_{r}$ only contains a regular self-adjoint operator
(namely $O(x^{\prime },x)^{\ast }=O(x,x^{\prime })).$

The states must be considered as linear functionals over the space $\mathcal{%
O}$ ($\mathcal{O}^{\prime}$ the dual of space $\mathcal{O}$):

\begin{equation}
\mathcal{O}^{\prime }\mathcal{=N}_{vhS}^{\prime }\mathcal{=N}_{S}^{\prime
}\oplus \mathcal{N}_{R}^{\prime }\subset \mathcal{N}^{\prime }  \label{m}
\end{equation}%
Therefore the states read as in eq. (\ref{c6}) or eq.(\ref{d2}). The set of
these generalized states is the convex set $\mathcal{S\subset O}^{\prime }$.

Now we can apply the projector of eq.(\ref{c10}) that in terms of the
algebra reads:

\begin{equation}
\Pi =\Pi ^{(0)}\oplus ...\oplus \Pi ^{(p)}:\mathcal{N}_{vhS}^{\prime
}\rightarrow \mathcal{N}_{R}^{\prime }  \label{proy}
\end{equation}

\textit{This is the simple trick that allows us to neglect the singularities
(i.e. the }$\delta (x-x^{\prime }))$ \textit{in a rigorous mathematical way
and to obtain correct physical results}. Essentially we have defined a new
dual space $\mathcal{O}^{\prime }$ (that contains the states $\rho $ without
divergences) that are adapted to solve our problem.

So, essentially we have substituted an \textquotedblright ad
hoc\textquotedblright\ counterterm procedure (or an ad hoc subtraction
procedure \cite{ref}) with a clear physical motivated theory.\textit{\ }%
These are the essential features of the proposed formalism, where the deltas
are absent.\footnote{%
This can also be considered as a way to multiply distributions (as in ref. 
\cite{Heep})}

\section{Summary}

Summarizing, the main idea of this work is that in the $p$ order in the
perturbation expansion of $\phi ^{4}$ theory, the state reads

\begin{equation}
\rho ^{(p)}=\rho _{ext}^{(p)}\underset{i=0}{\overset{p}{\otimes }}\rho
_{int}^{(i,p)}  \label{tr1}
\end{equation}%
and the observable reads

\begin{equation}
O=O_{ext}\underset{i=0}{\overset{p}{\otimes }}I_{int}^{(i)}  \label{tr2}
\end{equation}%
then

\begin{equation}
Tr(\rho ^{(p)}O)=Tr(\overline{\rho }_{ext}^{(p)}O_{ext})  \label{tr3}
\end{equation}%
where $\overline{\rho }_{ext}^{(p)}=Tr_{int}\rho ^{(p)}$ is the reduced
state. Because the state of eq.(\ref{tr1}) is a tensor product, then

\begin{equation}
Tr(\rho ^{(p)}O)=Tr(\rho _{ext}^{(p)}O_{ext})\underset{i=0}{\overset{p}{\Pi }%
}Tr(\rho _{int}^{(i,p)})  \label{tr4}
\end{equation}%
Finally, we can proceed with the sum in $p$

\begin{equation}
\overset{+\infty }{\underset{p=0}{\sum }}\frac{i^{p}}{p!}Tr(\rho ^{(p)}O)=%
\overset{+\infty }{\underset{p=0}{\sum }}\frac{i^{p}}{p!}Tr(\rho
_{ext}^{(p)}O_{ext})\underset{i=0}{\overset{p}{\Pi }}Tr(\rho _{int}^{(i,p)})
\label{tr5}
\end{equation}%
where $\underset{i=0}{\overset{p}{\Pi }}Tr(\rho _{int}^{(i,p)})$ is the
factor that contains the divergences. These divergences appear because the
internal quantum state contains diagonal functions multiplied by Dirac
deltas that cannot be avoid unless we assume that the diagonal functions are
zero, that can be obtained by a \textquotedblleft
projection\textquotedblright\ or by making a transformation on the diagonal
and non-diagonal functions.

From the point of view of the physics, the internal quantum state refers to
the internal vertices that appear in the perturbation expansion. The
particles that propagate to an internal vertice is called a virtual particle
because it can be off-shell, so there are not real and cannot be detected.
The mathematical formalism introduced in this work naturally consider these
virtual particles by assigning them a quantum state. But we cannot observe
these particles, so any relevant observable defined in the theory will be a
observable that acts on the external quantum state which refers to the
external particles. In terms of the mathematical formalism, this observable
will act as an identity in the internal quantum states of the virtual
particles. The consequence is that we can reduce the degrees of freedom of
the virtual particles with the result of a partial trace with respect to the
internal quantum system. This partial trace implies that we will integrate
over the degrees of freedom of the internal quantum state. This give us an
interpretation of this integration as a reduction of the degrees of freedom
of the theory. In the conventional interpretation of this integration
\textquotedblleft The integral $d^{4}z$ instructs us to sum over all points
where this process can ocurr. This is just the superposition principle of
quantum mechanics:\ when a process can happen in alternative ways, we add
the amplitudes for each possible way.\textquotedblright , (\cite{PS}, page
94). Then the fact that the reduction of the degrees of freedom results in a
divergent quantity comes from the fact that it allow the internal quantum
state to be singular by itself, in the sense that it can have a diagonal
function multiplied by a Dirac delta. The fact that the observable does not
look at the internal quantum state means that the diagonal function survives
and manifests itself in the mean value of that observable in the total
quantum state as a divergent quantity. So the projection procedure is to
take one virtual particle and eliminate his diagonal part.

Perhaps the most interesting of all this mathematical procedure developed to
treat $\phi ^{4}$ theory, which is a renormalizable theory, is that it could
be applied to non-renormalizable theories, such as $\phi ^{6}$. Apparently,
the procedure should not be different and for each correlation function one
can construct a quantum state that contains both an internal and external
part. Then we can construct a particular transformation or projection that
gives us a quantum state without a diagonal part. This would be the physical
contribution to the scattering process. In appendix C and D it is shown how
to apply the formalism introduced in Section 3 to the second order and first
order in the perturbation expansion of the self-energy of the electron in $%
QED$ and the self-energy of a scalar field with $\phi ^{6}$ self-interaction.

\section{Conclusions and prospects.}

The aim of the paper can be reassumed as follows. If, in order to explain
decoherence of quantum systems some procedures are allowed, then the same
procedures ought to be allowed to demonstrate the success of QFT. If we
accept this idea, the projection $\Pi $ and the choice of nice functions for
the set of observables and states are legitimate and then we could also
solve the main interpretative problems of QFT.

Of course it can be argued that these structures and properties are put
\textquotedblleft just by hand\textquotedblright . The answer is that all
mathematical structures and their properties (from the Galilei law of square
times to superstrings) are just choices made by physicists to explain nature
(and therefore also put by hand). The real art is to find the mathematical
structures to explain nature in the simplest way.

A lot of work must be done to transform this primitive idea into an
axiomatic based, mathematically rigorous, and finite QFT. But the main lines
of the picture have already been drawn.

It seems that these conclusions are in complete agreement with section 12.3
of \cite{W} and section 7.12 \cite{Folland}. In paper \cite{I} and in the
examples above we\ show in detail that our method is equivalent to usual
renormalization. These examples and the just quoted reference are enough to
foresee that this equivalence could be extended to more examples: So, may
be, our method could not only be applied to \textquotedblleft
renormalizable\textquotedblright\ theories with a finite number of
counterterms but also to \textquotedblright
non-renormalizable\textquotedblright\ theories with an infinite number of
arbitrary counterterms.

\section{Acknowledgment}

This paper was partially supported by grants of CONICET (Argentina National
Research Council), FONCYT (Argentina Agency for Science and Technology) and
the University of Buenos Aires.

\appendix

\section{Counting of ultraviolet divergences in $\frac{\protect\lambda }{l!}%
\protect\phi ^{l}$ theory}

Let us consider a pure scalar field theory with an interaction term $\frac{%
\lambda }{l!}\phi ^{l}$. Let $r_{I}$ be the number of internal propagators
(propagators that are not connected to external points) and $p$ the number
of vertices. Then, the number of loops in a Feynman diagram reads (see \cite%
{PS}, pag. 321):

\begin{equation}
L=r_{I}-p+1  \label{co1}
\end{equation}

The number of internal propagators $r_{I}$ can be written in terms of the
number of external points$\,n$, the number of vertices $p$ and $l$. The
total number of propagators $r$ in a Feynman diagram is:

\begin{equation}
r=r_{I}+r_{E}  \label{co2}
\end{equation}%
where $r_{E}$ is the number of external propagators or external lines which
coincide with the number of external points.\footnote{%
The contribution to the generating functional comes from the connected
Feynman diagrams. This means that each external point must be connected to a
vertex. If there are $n$ external points then there will be $n$ external
lines.} In turn, if the correlation function has $n$ external fields and $%
l\cdot p$ internal fields, then the total number of propagators $r~$reads:

\begin{equation}
r=\frac{n+l\cdot p}{2}  \label{co3}
\end{equation}%
Then, replacing eq.(\ref{co3}) in eq.(\ref{co2}) we have:

\begin{equation}
r_{I}=r-r_{E}=\frac{l\cdot p}{2}-\frac{n}{2}  \label{co4}
\end{equation}%
Replacing eq.(\ref{co4})\ in eq.(\ref{co1}) we finally have:

\begin{equation}
L(l,p,n)=p\left( \frac{l-2}{2}\right) -\frac{n}{2}+1  \label{co5}
\end{equation}%
Each loop contributes with a term proportional to $\frac{1}{\epsilon }$ plus
a finite term (see \cite{critical}, page 103-130 and \cite{ashok}, page
686). Because the loops are multiplied together in a Feynman diagram of a $%
\frac{\lambda }{l}\phi ^{l}$ theory with $n$ external points and $p$
vertices, we obtain the following divergent term:

\begin{equation}
\Omega _{p}^{(p,k)}(l,p,n)=\underset{k=0}{\overset{L(l,p,n)-1}{\dsum }}\frac{%
\beta _{L-k}^{(p,k)}}{\epsilon ^{L-k}}  \label{co6}
\end{equation}%
For example, for $l=4$, $n=2$ we have%
\begin{equation}
\Omega _{p}^{(p,k)}(4,p,2)=\underset{n=0}{\overset{p}{\dsum }}\frac{\beta
_{p-n}^{(p,k)}}{\epsilon ^{p-n}}=\frac{\beta _{p}^{(p,k)}}{\epsilon ^{p}}%
+...+\beta _{0}^{(p,k)}  \label{co7}
\end{equation}%
which coincides with the divergent structure of eq.(\ref{d5.1}).

\section{Two-point correlation function of the self-interacting $\protect%
\phi ^{4}$ theory}

Let us consider a self-interacting scalar field with a $\frac{\lambda }{4!}%
\phi ^{4}\,$interaction. The two-points connected correlation function,
which represent the propagator in the interacting theory reads

\begin{equation}
\left\langle \Omega \left\vert \phi (x_{1})\phi (x_{2})\right\vert \Omega
\right\rangle =\overset{+\infty }{\underset{p=0}{\sum }}\frac{1}{p!}(\frac{%
i\lambda }{4!})^{p}\dint \left\langle \Omega _{0}\left\vert T\phi
_{0}(x_{1})\phi _{0}(x_{2})\phi ^{4}(y_{1})...\phi ^{4}(y_{p})\right\vert
\Omega _{0}\right\rangle d^{4}y_{1}...d^{4}y_{p}  \label{pro1}
\end{equation}%
Resolving the correlation function inside each integral and in each
perturbation term we have:

\begin{eqnarray}
\left\langle \Omega \left\vert \phi (x_{1})\phi (x_{2})\right\vert \Omega
\right\rangle &=&\left\langle \Omega _{0}\left\vert T\phi _{0}(x_{1})\phi
_{0}(x_{2})\right\vert \Omega _{0}\right\rangle +\frac{i\lambda }{4!}\dint
\left\langle \Omega _{0}\left\vert T\phi _{0}(x_{1})\phi _{0}(x_{2})\phi
^{4}(y_{1})\right\vert \Omega _{0}\right\rangle d^{4}y_{1}+  \label{pro6} \\
&&\frac{1}{2!}(\frac{i\lambda }{4!})^{2}\dint \left\langle \Omega
_{0}\left\vert T\phi _{0}(x_{1})\phi _{0}(x_{2})\phi ^{4}(y_{1})\phi
^{4}(y_{2})\right\vert \Omega _{0}\right\rangle d^{4}y_{1}d^{4}y_{2}+... 
\notag
\end{eqnarray}%
It can be shown that using dimensional regularization, each term in the
perturbation can be written as

\begin{equation}
\left\langle \Omega _{0}\left\vert T\phi _{0}(x_{1})\phi
_{0}(x_{2})\right\vert \Omega _{0}\right\rangle =\dint \frac{d^{4}k}{(2\pi
)^{4}}\frac{e^{-ik(x_{1}-x_{2})}}{k^{2}-m_{0}^{2}}  \label{q1}
\end{equation}%
where $k$ is the external momentum. For the first order in the perturbation
we have

\begin{equation}
\frac{i\lambda }{4!}\dint \left\langle \Omega _{0}\left\vert T\phi
_{0}(x_{1})\phi _{0}(x_{2})\mathcal{L}(y_{1})\right\vert \Omega
_{0}\right\rangle d^{4}y=\frac{i\lambda }{4!}\dint \frac{d^{4}k}{(2\pi )^{4}}%
\frac{e^{-ik(x_{1}-x_{2})}}{(k^{2}-m_{0}^{2})^{2}}\left( \beta
_{0}^{(1,0)}+\beta _{1}^{(1,0)}\frac{1}{\epsilon }\right)  \label{q2}
\end{equation}%
where $\beta _{0}^{(1,0)}$and $\beta _{1}^{(1,0)}$ are some constants which
are functions of $\mu $, a mass factor introduced by changing the coupling
constant as $\lambda $ $\rightarrow $ $\lambda (\mu ^{2})^{-\epsilon }$ to
keep it dimensioneless, $\epsilon =d-4$, where $d$ is the dimension of
space-time and $m_{0}$ is the bare mass. The first superscript in the
constants refers to the order in the perturbation and the second one to the
power of the propagator minus one. The contribution $\beta
_{0}^{(1,0)}+\beta _{1}^{(1,0)}\frac{1}{\epsilon }$ for the first order
comes from the tadpole diagram, where a $\Delta (0)$ appears. If we use
dimensional regularization we find that

\begin{eqnarray}
(\mu ^{2})^{-\epsilon }\Delta (0) &=&\frac{m_{0}^{2}}{(4\pi )^{2}}\left[
1-\epsilon \ln (\frac{4\pi \mu }{m_{0}^{2}}^{2})+O(d-4)\right] \left[ \frac{2%
}{\epsilon }+\Psi (2)+O(\epsilon )\right] =  \label{q2.1} \\
&&\frac{m_{0}^{2}}{(4\pi )^{2}}\left[ \Psi (2)-2\ln (\frac{m_{0}^{2}}{4\pi
\mu ^{2}})+\frac{2}{\epsilon }+O(\epsilon )\right]  \notag
\end{eqnarray}%
where $\Psi (2)=1-\gamma $ where $\gamma $ is the Euler-Mascheroni constant,
then $\beta _{0}^{(1,0)}=\frac{m_{0}^{2}}{(4\pi )^{2}}\left[ \Psi (2)-2\ln (%
\frac{m_{0}^{2}}{4\pi \mu ^{2}})\right] $ and $\beta _{1}^{(1,0)}=\frac{%
2m_{0}^{2}}{(4\pi )^{2}}$.

For the second order in the perturbation theory we have (see \cite{Ramond},
page 119-125):

\begin{gather}
(\frac{i\lambda }{4!})^{2}(\mu ^{2})^{2(4-d)}\dint \left\langle \Omega
_{0}\left\vert T\phi _{0}(x_{1})\phi _{0}(x_{2})\mathcal{L}(y_{1})\mathcal{L}%
(y_{2})\right\vert \Omega _{0}\right\rangle d^{4}y_{1}d^{4}y_{2}=  \label{q3}
\\
(\frac{i\lambda }{4!})^{2}\dint \frac{d^{4}k}{(2\pi )^{4}}\frac{%
e^{-ik(x_{1}-x_{2})}}{(k^{2}-m_{0}^{2})^{2}}\left( \beta _{0}^{(2,0)}+\beta
_{1}^{(2,0)}\frac{1}{\epsilon }+\beta _{2}^{(2,0)}\frac{1}{\epsilon ^{2}}%
\right) +  \notag \\
(\frac{i\lambda }{4!})^{2}\dint \frac{d^{4}k}{(2\pi )^{4}}\frac{%
e^{-ik(x_{1}-x_{2})}}{(k^{2}-m_{0}^{2})^{3}}\left( \beta _{0}^{(2,1)}+\beta
_{1}^{(2,1)}\frac{1}{\epsilon }+\beta _{2}^{(2,1)}\frac{1}{\epsilon ^{2}}%
\right)  \notag
\end{gather}

In this case, we have two differents powers in the external propagator. The
reason is that in the second order in the perturbation theory, a Feynmann
diagram will be irreducible and the other not. The Feynmann diagram that is
not irreducible is given by two loops connected each other by a propagator
and each of them connected to the external lines. As we can see, the
perturbation expansion is also an expansion in the number of loops. When we
proceed with dimensional regularization, each loop contributes with $a+\frac{%
b}{\epsilon }$.

We can continue with the following orders and finally obtain the following
result when $p>1$:

\begin{equation}
\dint \left\langle \Omega _{0}\left\vert T\phi _{0}(x_{1})\phi _{0}(x_{2})%
\mathcal{L}(y_{1})...\mathcal{L}(y_{p})\right\vert \Omega _{0}\right\rangle
d^{4}y_{1}...d^{4}y_{p}=\underset{l=0}{\overset{p}{\sum }}\dint \frac{d^{4}k%
}{(2\pi )^{4}}\frac{e^{-ik(x_{1}-x_{2})}}{(k^{2}-m_{0}^{2})^{l+2}}\underset{%
j=0}{\overset{p}{\sum }}\beta _{j}^{(p,l)}\frac{1}{\epsilon ^{j}}  \label{q4}
\end{equation}%
Now we can proceed with the sum in $p\,$as eq.(\ref{pro1}) indicates:

\begin{equation}
\left\langle \Omega \left\vert \phi (x_{1})\phi (x_{2})\right\vert \Omega
\right\rangle =\dint \frac{d^{4}k}{(2\pi )^{4}}\frac{e^{-ik(x_{1}-x_{2})}}{%
k^{2}-m_{0}^{2}}+\overset{+\infty }{\underset{p=1}{\sum }}\frac{1}{p!}(\frac{%
i\lambda }{4!})^{p}\underset{l=0}{\overset{p}{\sum }}\dint \frac{d^{4}k}{%
(2\pi )^{4}}\frac{e^{-ik(x_{1}-x_{2})}}{(k^{2}-m_{0}^{2})^{l+2}}\underset{j=0%
}{\overset{p}{\sum }}\beta _{j}^{(p,l)}\frac{1}{\epsilon ^{j}}  \label{q5}
\end{equation}%
Rearranging the sum, eq.(\ref{q5}) can be written as:

\begin{equation}
\left\langle \Omega \left\vert \phi (x_{1})\phi (x_{2})\right\vert \Omega
\right\rangle =\dint \frac{d^{4}k}{(2\pi )^{4}}\frac{e^{-ik(x_{1}-x_{2})}}{%
k^{2}-m_{0}^{2}}+\overset{+\infty }{\underset{s=0}{\sum }}\dint \frac{d^{4}k%
}{(2\pi )^{4}}\frac{e^{-ik(x_{1}-x_{2})}}{(k^{2}-m_{0}^{2})^{2+s}}\overset{%
+\infty }{\underset{n=0}{\sum }}\overset{+\infty }{\underset{j=1}{\sum }}(%
\frac{i\lambda }{4!})^{j}\beta _{n}^{(j,s)}\frac{1}{\epsilon ^{n}}
\label{q6}
\end{equation}%
Now we can put $x_{2}=0$ and make the Fourier transformation:

\begin{equation}
\dint \frac{d^{4}p}{(2\pi )^{4}}e^{-ipx_{1}}\left\langle \Omega \left\vert
\phi (x_{1})\phi (x_{0})\right\vert \Omega \right\rangle =\frac{1}{%
p^{2}-m_{0}^{2}}+\overset{+\infty }{\underset{s=0}{\sum }}\frac{1}{%
(p^{2}-m_{0}^{2})^{2+s}}\left( \overset{+\infty }{\underset{j=1}{\sum }}(%
\frac{i\lambda }{4!})^{j}\beta _{0}^{(j,s)}+\overset{+\infty }{\underset{n=1}%
{\sum }}\overset{+\infty }{\underset{j=1}{\sum }}(\frac{i\lambda }{4!}%
)^{j}\beta _{n}^{(j,s)}\frac{1}{(d-4)^{n}}\right)  \label{q7}
\end{equation}%
where we have separated the terms with $\frac{1}{\epsilon ^{0}}$.

With the purpose of neglecting the terms that depend on the space-time
dimension $d$, we can make the following transformation

\begin{equation}
\beta _{0}^{(j,s)}=\overline{\beta }_{0}^{(j,s)}-\overset{+\infty }{\underset%
{n=1}{\sum }}\alpha _{n}^{(j,s)}\frac{1}{\epsilon ^{n}}  \label{q8}
\end{equation}%
where $\alpha _{n}^{(j,s)}$ are some constants that will cancel the
contributions of\ $\gamma _{n}^{(j,s)}$ in eq.(\ref{q7}). Then, the term
inside the bracket in the r.h.s of eq.(\ref{q7}) reads

\begin{equation}
\overset{+\infty }{\underset{j=1}{\sum }}(\frac{i\lambda }{4!})^{j}\overline{%
\beta }_{0}^{(j,s)}-\overset{+\infty }{\underset{j=1}{\sum }}\overset{%
+\infty }{\underset{n=1}{\sum }}(\frac{i\lambda }{4!})^{j}\alpha _{n}^{(j,s)}%
\frac{1}{\epsilon ^{n}}+\overset{+\infty }{\underset{n=1}{\sum }}\overset{%
+\infty }{\underset{j=1}{\sum }}(\frac{i\lambda }{4!})^{j}\beta _{n}^{(j,s)}%
\frac{1}{\epsilon ^{n}}=0  \label{q9}
\end{equation}%
which implies that

\begin{equation}
\alpha _{n}^{(j,s)}-\beta _{n}^{(j,s)}=0  \label{q10}
\end{equation}%
then

\begin{equation}
\dint \frac{d^{4}p}{(2\pi )^{4}}e^{-ipx_{1}}\left\langle \Omega \left\vert
\phi (x_{1})\phi (x_{0})\right\vert \Omega \right\rangle =\frac{1}{%
p^{2}-m_{0}^{2}}+\overset{+\infty }{\underset{s=0}{\sum }}\frac{1}{%
(p^{2}-m_{0}^{2})^{2+s}}M(s)  \label{q11}
\end{equation}%
where

\begin{equation}
M(s)=\overset{+\infty }{\underset{j=1}{\sum }}(\frac{i\lambda }{4!})^{j}%
\overline{\beta }_{0}^{(j,s)}  \label{q12}
\end{equation}%
is the finite contribution to the propagator of the self-interacting scalar
theory. Now, each of this terms $M(s)$ depends on $s$ which is the power of
the external propagator. We know that $M(0)$ is the one-particle irreducible
diagram and the following terms $M(s)$ with $s>1$ are the product of this $%
M(0)$:

\begin{equation}
M(s)=\left[ M(0)\right] ^{s}  \label{q13}
\end{equation}%
Introducing this last result in eq.(\ref{q11}) we have

\begin{equation}
\dint \frac{d^{4}p}{(2\pi )^{4}}e^{-ipx_{1}}\left\langle \Omega \left\vert
\phi (x_{1})\phi (x_{0})\right\vert \Omega \right\rangle =\frac{1}{%
p^{2}-m_{0}^{2}}\overset{+\infty }{\underset{s=0}{\sum }}(\frac{M(0)}{%
p^{2}-m_{0}^{2}})^{s}=\frac{1}{p^{2}-m_{0}^{2}}\frac{1}{1-\frac{M(0)}{%
p^{2}-m_{0}^{2}}}=\frac{1}{p^{2}-m_{0}^{2}-M(0)}  \label{q14}
\end{equation}%
which is our desired result. The propagator of the self-interacting scalar
theory has a pole which is shifted away by

\begin{equation}
m^{2}=m_{0}^{2}+M(0)=m_{0}^{2}+\overset{+\infty }{\underset{j=1}{\sum }}(%
\frac{i\lambda }{4!})^{j}\beta _{0}^{(j,0)}  \label{q15}
\end{equation}%
If we do not make the transformation of eq.(\ref{q8}), then the shift in the
mass\footnote{%
Of course, to find the following result we must do the sum $\underset{n=0}{%
\sum }x^{n}=\frac{1}{1-x}$ but in this case $x$ is not less than one.} would
be:

\begin{equation}
m^{2}=m_{0}^{2}+\overset{+\infty }{\underset{j=1}{\sum }}(\frac{i\lambda }{4!%
})^{j}\beta _{0}^{(j,0)}+\overset{+\infty }{\underset{n=1}{\sum }}\overset{%
+\infty }{\underset{j=1}{\sum }}(\frac{i\lambda }{4!})^{j}\beta _{n}^{(j,0)}%
\frac{1}{\epsilon ^{n}}  \label{q16}
\end{equation}%
which is identical to eq.(2.3a) of \cite{thooft}. Eq.(\ref{q15}) and eq.(\ref%
{q16}) give the renormalization group, because $\beta _{0}^{(j,0)}$ depends
on the mass factor $\mu $, $\epsilon $, $m_{0}$ and $\lambda _{0}$, so in
the general case, the unrenormalized two-point correlation function or Green
function $\Gamma _{0}^{2}$depends on $\mu $, $\epsilon $, $m_{0}$ and $%
\lambda _{0}$ and the renormalized $\Gamma ^{2}$ two-point correlation
function only depends on $m$, $\lambda $ and $\epsilon $.

\section{Second order in $QED$ for the self-electron energy}

We can apply the same formalism to $QED$. We can define the following
quantum state and observable:

\begin{equation}
\rho ^{(2)}=\dint\limits_{{}}^{{}}\rho _{ext}^{(2)}(x_{1},x_{2})\left( \rho
_{D}^{(1)}(y_{1})\delta (y_{1}-w_{1})+\rho _{ND}^{(1)}(y_{1},w_{1})\right)
\left\vert x_{1},y_{1}\right\rangle \left\langle x_{2},w_{1}\right\vert
d^{4}x_{1}d^{4}x_{2}d^{4}y_{1}d^{4}w_{1}  \label{sec1}
\end{equation}

\begin{equation}
O^{(2)}=\dint\limits_{{}}^{{}}O_{ext}(x_{1},x_{2})\delta
(y_{1}-w_{1})\left\vert x_{1},y_{1}\right\rangle \left\langle
x_{2},w_{1}\right\vert d^{4}x_{1}d^{4}x_{2}d^{4}y_{1}d^{4}w_{1}  \label{sec2}
\end{equation}%
the mean value $Tr(\rho ^{(2)}O^{(2)})$ reads%
\begin{equation}
Tr(\rho ^{(2)}O^{(2)})=\left( \rho _{D}^{(1)}\delta (0)+\rho
_{ND}^{(1)}\right) Tr(\rho _{ext}^{(2)}O_{ext})  \label{sec3}
\end{equation}%
where

\begin{equation}
Tr(\rho _{ext}^{(2)}O_{ext})=\dint\limits_{{}}^{{}}\rho
_{ext}^{(2)}(x_{1},x_{2})O_{ext}(x_{1},x_{2})d^{4}x_{1}d^{4}x_{2}
\label{sec4}
\end{equation}%
and

\begin{equation}
\rho _{D}^{(1)}=\dint\limits_{{}}^{{}}\rho _{D}^{(1)}(y_{1})d^{4}y_{1}\text{
\ \ \ \ \ \ \ \ \ \ \ \ }\rho _{ND}^{(1)}=\dint\limits_{{}}^{{}}\rho
_{ND}^{(1)}(y_{1},y_{1})d^{4}y_{1}  \label{sec5}
\end{equation}%
If we replace

\begin{gather}
\rho _{ext}^{(2)}(x_{1},x_{2})=\dint \frac{d^{4}p}{(2\pi )^{4}}\frac{i(\NEG%
{p}_{\alpha }+m)e^{-ip(x-y)}}{p^{2}-m^{2}}\text{ \ \ \ \ \ \ }%
O_{ext}(x_{1},x_{2})=J(x_{1})J(x_{2})  \label{sec6} \\
\text{\ }\rho _{D}^{(1)}=\frac{-\NEG{p}+4m}{8\pi ^{2}}\text{ \ \ \ \ \ \ \ \
\ \ \ \ \ \ \ \ \ \ \ \ \ \ \ \ \ \ \ }\delta (0)=\underset{\epsilon
\rightarrow 0}{\lim }\frac{1}{\epsilon }  \notag \\
\text{\ }\rho _{ND}^{(1)}=\frac{e^{2}}{8\pi ^{2}}\left[ \frac{1}{2}\NEG%
{p}(1+\gamma )+m(1+2\gamma )+\dint\limits_{0}^{1}dx\left[ \NEG{p}(1-x)+2m%
\right] \ln \left( \frac{p^{2}x(1-x)+m^{2}x}{4\pi \mu ^{2}}\right) \right] 
\notag
\end{gather}%
then eq.(\ref{sec3}) is equal to eq.(8.2.20) of \cite{Ramond}. In this
sense, $\rho _{ND}^{(1)}$ is the finite contribution to the self-energy of
the electron. In the projection procedure introduced in this work, the
finite contribution would directly be given by $\rho _{ND}^{(1)}$.

\section{First order in $\protect\phi ^{6}$}

The formalism developed in this work allows us to apply it to theories that
are in principle not renormalizable, as $\phi ^{6}$. In this appendix it is
shown the first order in the perturbation $\phi ^{6}$ theory. Nevertheless,
this non-renormalizable theory will be developed in details in future works.

We can define a state and an observable at the first order as

\begin{gather}
\rho ^{(1)}=\dint\limits_{{}}^{{}}\rho _{ext}^{(1)}(x_{1},x_{2})\left( \rho
_{D}^{(1)}(y_{1})\delta (y_{1}-w_{1})+\rho _{ND}^{(1)}(y_{1},w_{1})\right)
\left( \rho _{D}^{(2)}(y_{2})\delta (y_{2}-w_{2})+\rho
_{ND}^{(2)}(y_{2},w_{2})\right)  \label{gaton1} \\
\left\vert x_{1},y_{1},y_{2}\right\rangle \left\langle
x_{2},w_{1},w_{2}\right\vert
d^{4}x_{1}d^{4}x_{2}d^{4}y_{1}d^{4}w_{1}d^{4}y_{2}d^{4}w_{2}  \notag
\end{gather}

\begin{equation}
O^{(1)}=\dint\limits_{{}}^{{}}O_{ext}(x_{1},x_{2})\delta (y_{1}-w_{1})\delta
(y_{2}-w_{2})\left\vert x_{1},y_{1},y_{2}\right\rangle \left\langle
x_{2},w_{1},w_{2}\right\vert
d^{4}x_{1}d^{4}x_{2}d^{4}y_{1}d^{4}w_{1}d^{4}y_{2}d^{4}w_{2}  \label{gaton2}
\end{equation}%
then, the mean value $Tr(\rho ^{(1)}O^{(1)})\ $reads

\begin{equation}
Tr(\rho ^{(1)}O^{(1)})=\left( \rho _{D}^{(1)}\rho _{D}^{(2)}\left[ \delta (0)%
\right] ^{2}+(\rho _{D}^{(1)}\rho _{ND}^{(2)}+\rho _{ND}^{(1)}\rho
_{D}^{(2)})\delta (0)+\rho _{ND}^{(1)}\rho _{ND}^{(2)}\right) Tr(\rho
_{ext}^{(1)}\ O_{ext})  \label{gaton3}
\end{equation}%
where

\begin{equation}
Tr(\rho _{ext}^{(1)}O_{ext})=\dint\limits_{{}}^{{}}\rho
_{ext}^{(1)}(x_{1},x_{2})O_{ext}(x_{1},x_{2})d^{4}x_{1}d^{4}x_{2}
\label{gaton4}
\end{equation}

\begin{equation}
\rho _{D}^{(i)}=\dint\limits_{{}}^{{}}\rho _{D}^{(i)}(y_{i})d^{4}y_{i}\text{
\ \ \ \ \ \ \ \ \ \ \ \ }\rho _{ND}^{(i)}=\dint\limits_{{}}^{{}}\rho
_{ND}^{(i)}(y_{i},y_{i})d^{4}y_{i}  \label{gaton5}
\end{equation}%
The first order in the perturbation expansion in $\phi ^{6}$ theory reads

\begin{gather}
\dint\limits_{{}}^{{}}\left\langle \Omega _{0}\left\vert \phi (x_{1})\phi
(x_{2})\phi ^{6}(y_{1})\right\vert \Omega _{0}\right\rangle
J(x_{1})J(x_{2})d^{4}y_{1}d^{4}x_{1}d^{4}x_{2}=  \label{gaton6} \\
\left[ \Delta (0)\right] ^{2}\dint \Delta (x_{1}-y_{1})\Delta
(x_{2}-y_{1})J(x_{1})J(x_{2})d^{4}y_{1}d^{4}x_{1}d^{4}x_{2}  \notag
\end{gather}%
The integral in the internal coordinate $y_{1}$ reads

\begin{equation}
\dint \Delta (x_{1}-y_{1})\Delta
(x_{2}-y_{1})d^{4}y_{1}=\dint\limits_{{}}^{{}}\frac{d^{4}p}{(2\pi )^{4}}%
\frac{e^{-ip(x_{1}-x_{2})}}{(p^{2}-m^{2})^{2}}  \label{gaton7}
\end{equation}%
and

\begin{equation}
\left[ \Delta (0)\right] ^{2}=\left( \frac{\alpha _{2}}{\epsilon ^{2}}+\frac{%
\alpha _{1}}{\epsilon }+\alpha _{0}\right)  \label{gaton8}
\end{equation}%
where $\alpha _{0}=\left[ \Psi (2)\right] ^{2}$, $\alpha _{1}=4\Psi (2)$ and 
$\alpha _{2}=4$, see eq.(\ref{q2.1}). Then, eq.(\ref{gaton6}) finally reads

\begin{equation}
\dint\limits_{{}}^{{}}\left\langle \Omega _{0}\left\vert \phi (x_{1})\phi
(x_{2})\phi ^{6}(y_{1})\right\vert \Omega _{0}\right\rangle
J(x_{1})J(x_{2})d^{4}y_{1}d^{4}x_{1}d^{4}x_{2}=\left( \frac{\alpha _{2}}{%
\epsilon ^{2}}+\frac{\alpha _{1}}{\epsilon }+\alpha _{0}\right)
\dint\limits_{{}}^{{}}\frac{d^{4}p}{(2\pi )^{4}}\frac{e^{-ip(x_{1}-x_{2})}}{%
(p^{2}-m^{2})^{2}}J(x_{1})J(x_{2})d^{4}x_{1}d^{4}x_{2}  \label{gaton9}
\end{equation}%
This last equation is similar to eq.(\ref{gaton3}), in fact, if we replace

\begin{gather}
\rho _{ext}^{(1)}(x_{1},x_{2})=\dint\limits_{{}}^{{}}\frac{d^{4}p}{(2\pi
)^{4}}\frac{e^{-ip(x_{1}-x_{2})}}{(p^{2}-m^{2})^{2}}\text{ \ \ \ \ \ \ \ }%
O_{ext}(x_{1},x_{2})=J(x_{1})J(x_{2})  \label{gaton10} \\
\rho _{D}^{(1)}\rho _{D}^{(2)}=\alpha _{2}\text{ \ \ \ \ }\rho
_{D}^{(1)}\rho _{ND}^{(2)}+\rho _{ND}^{(1)}\rho _{D}^{(2)}=\alpha _{1}\text{
\ \ \ \ \ \ }\rho _{ND}^{(1)}\rho _{ND}^{(2)}=\alpha _{0}  \notag \\
\left[ \delta (0)\right] ^{n}=\underset{\epsilon \rightarrow 0}{\lim }\frac{1%
}{\epsilon ^{n}}  \notag
\end{gather}%
Eq.(\ref{gaton9}) and eq.(\ref{gaton3}) are identical. From this point of
view, the projection procedure would give the finite contribution that only
depends on the non-diagonal states~$\rho _{ND}^{(i)}$.

\end{document}